\documentstyle[epsf]{mn}
\def\spose#1{\hbox to 0pt{#1\hss}}
\def\lta{\mathrel{\spose{\lower 3pt\hbox{$\sim$}}
    \raise 2.0pt\hbox{$<$}}}
\def\gta{\mathrel{\spose{\lower 3pt\hbox{$\sim$}}
    \raise 2.0pt\hbox{$>$}}}
\title[The curvature condition for self-consistent scale-free galaxies]
{The curvature condition for self-consistent scale-free galaxies}
\author[M. A. Jalali and P. T. de Zeeuw]
	{Mir Abbas Jalali$^{1,2}$\thanks{jalali@iasbs.ac.ir}
	and P. Tim de Zeeuw$^3$\thanks{tim@strw.leidenuniv.nl} \\ 
	$^1$Institute for Advanced Studies in Basic Sciences,
	P.O. Box 45195-159, Zanjan, IRAN \\
	$^2$Aerospace Research Institute, 
	P.O. Box 15875-3885, Tehran, IRAN \\ 
	$^3$Sterrewacht Leiden, Postbus 9513, 2300 RA Leiden,
	The Netherlands}  
  
\begin{document}  
\label{firstpage}  
\maketitle  
  
\begin{abstract}   
We modify the curvature condition for the existence of self-consistent 
scale-free discs, introduced by Zhao, Carollo \& de Zeeuw. We survey 
the parameter space of the power-law discs, and show that the modified 
curvature condition is in harmony with the results of Schwarzschild's 
numerical orbit superposition method. We study the orbital structure 
of the power-law discs, and find a correlation between the population 
of centrophobic banana orbits and the non-self-consistency index. We 
apply the curvature condition to other families of scale-free 
elongated discs and find that it rules out a large range of power-law 
slopes and axis ratios. We generalize the condition, and apply it, to 
three-dimensional scale-free axisymmetric galaxy models.\looseness=-2 
\end{abstract}  

\begin{keywords}  
stellar dynamics -- galaxies: kinematics and dynamics -- galaxies: 
structure -- methods: analytical. 
\end{keywords}  
  
\section{Introduction}  
\label{sec:introduction} 
 
A key problem in stellar dynamics is the determination of the range of 
axis ratios and density profiles for which triaxial galaxies can be in 
dynamical equilibrium (see, e.g., reviews by de Zeeuw 1996; Merritt 
1999). Schwarzschild (1979, 1982) attacked this issue with his linear 
programming method, and constructed self-consistent triaxial galaxy 
models with constant density cores by numerical superposition of 
individual stellar orbits. Similar models with separable potentials 
were constructed by Statler (1987) and Hunter \& de Zeeuw (1992). 
Gerhard \& Binney (1985) showed that the orbital structure in triaxial 
galaxies with central density cusps differs from those with cores. In 
particular, the box orbits are replaced by a host of minor orbit 
families and chaotic orbits (cf.\ Miralda--Escud\'e \& Schwarzschild 
1989; Lees \& Schwarzschild 1992). Based on this, Gerhard (1986) 
already suggested that cusped triaxial galaxies would evolve towards 
axisymmetry. Schwarzschild (1993) showed that moderately flattened 
scale-free triaxial models with an $r^{-2}$ density law could be 
constructed numerically, but found that these models evolved on long 
time scales. Similar results were obtained for non-scale-free models 
with steep central cusps (e.g., Merritt \& Fridman 1996; Merritt 1997; 
Siopis 1998).\looseness=-2 

Although the general problem of the existence of self-consistent 
triaxial, cuspy galaxies has not been solved, valuable steps have been 
taken in the study of planar models that inherit many properties of 
the three-dimensional triaxial systems. Kuijken (1993, hereafter K93) 
applied Schwarzschild's numerical construction method to elliptic 
discs with logarithmic potentials, and found that self-consistent 
models are ruled out for axial ratios $\lta 0.8$. He traced the 
non-existence to the properties of the chaotic orbits, in line with 
the three-dimensional results of Schwarzschild (1993). Sridhar \& 
Touma (1997a, hereafter STa) introduced a class of non-axisymmetric 
discs whose potentials are of St\"ackel form in parabolic 
coordinates. These discs admit an exact second integral of motion and 
host a continuous family of centrophobic banana orbits. Syer \& Zhao 
(1998, hereafter SZ98) showed by numerical means that none of the STa 
models can be self-consistent.\looseness=-2 
 
Zhao, Carollo \& de Zeeuw (1999, hereafter ZCZ) developed a simple 
analytical tool for studying the self-consistency of scale-free 
discs. They compared the curvature of the model density with that of 
individual orbits near the major axis, and showed that this leads to a 
necessary (but not sufficient) condition for self-consistency which 
can be evaluated easily. They analysed a family of elliptic discs with 
power-law densities for a range of slopes, and showed that the 
curvature condition produces results in harmony with those of K93 and 
SZ98. We show here that while the ZCZ approach is sound, their 
condition for self-consistency contains an unfortunate error, which 
leads to incorrect results for surface density distributions that do 
not fall off as $1/R$. We present the correct condition for 
two-dimensional scale-free models, and compare the results with models 
constructed by means of Schwarzschild's method. This not only 
validates the curvature condition, but also provides insight in the 
proper way to apply Schwarzschild's method. 
 
This paper is organized as follows. In \S\ref{sec:scale-free-discs} we
introduce scale-free elongated discs, describe how self-consistent
discs can be constructed with Schwarzschild's method, and rederive and
correct the curvature condition of ZCZ. In \S\ref{sec:power-law-discs}
we study the parameter ranges of self-consistent power-law discs, and
compare the curvature results with the numerical solutions. In
\S\ref{sec:other-discs} we apply the curvature condition to three
other families of elongated discs, two of which are new, and confirm
the results of K93 and SZ98. We extend the curvature condition to
three-dimensional scale-free axisymmetric models in
\S\ref{sec:curvature-three}, and apply it in \S\ref{sec:app-oblate}.
We summarize our conclusions in \S\ref{sec:discussion}. We will study
the triaxial case in a future paper.\looseness=-2

\section{Scale-free elongated discs} 
\label{sec:scale-free-discs} 
 
\subsection{Surface-density, potential, and scaling} 
\label{sec:pot-den} 
 
We consider razor-thin power-law discs with a surface density given by 
\begin{equation} 
\Sigma(R, \phi) = \Sigma_0 R^{\alpha-1} S(\phi),  
\label{eq:surf-density} 
\end{equation} 
and a corresponding gravitational potential  
\begin{eqnarray}  
V(R, \phi) \! &=& \left\{  
\begin{array}{ll} V_0 R^\alpha P(\phi),  &\alpha\not=0, \\ 
                  \null & \null \\ 
                  V_0 [2\ln R + P(\phi)], & \alpha=0,  \\ 
\end{array}\right. 
\label{eq:potential}  
\end{eqnarray}  
where $(R, \phi)$ are the usual polar coordinates, and the functions 
$S(\phi)$ and $P(\phi)$ are related by Poisson's equation. We are 
interested in models which are bisymmetric, and hence restrict 
ourselves to $0 \leq \phi \leq \pi/2$, where $\phi=0$ corresponds to 
the long axis. 
 
Since the potential and surface density functions of our discs are 
scale-free, the orbits at different energies are related by simple 
scaling factors in length and time (e.g., Richstone 1980, hereafter 
R80). To find these factors, we use the equations of motion: 
\begin{equation}  
\ddot R -R \dot \phi^2 = -\frac {\partial V}{\partial R}, \qquad   
R \ddot \phi +2 \dot R \dot \phi = -\frac 1R   
\frac {\partial V}{\partial \phi}.  
\label{eq:eqs-motion}  
\end{equation}  
For a potential of the form (\ref{eq:potential}) these equations are 
invariant under the scaling transformations 
\begin{equation}  
R= k \tilde R, \qquad t= k^{1-\frac{\alpha}{2}}\, \tau,  
\label{eq:scaling}  
\end{equation}  
where $\tau$ and $\tilde R$ are the scaled time and radius, 
respectively. This means that if a star with the coordinates 
$[R(t),\phi (t)]$ spends a time $\Delta t$ in the angular sector 
$\Delta \phi$, there exists another star with the coordinates $[\tilde 
R(\tau),\phi (\tau)]$ on the scaled orbit that spends the time 
\begin{equation}  
\Delta \tau =\left (\frac {R}{\tilde R} \right )^{\frac   
{\alpha}{2}-1} \Delta t,  
\label{eq:scaled-time}  
\end{equation}  
in the same angular sector. It follows that the velocity vector $(v_R, 
v_\phi)$ scales as $(k^{\frac{\alpha}{2}} v_R, k^{\frac{\alpha}{2}} 
v_\phi)$.

\subsection{Schwarzschild's method}  
\label{sec:schwarzschild} 
  
The numerical construction of the self-consistent discs defined by 
eqs~(\ref{eq:surf-density}) and (\ref{eq:potential}) is simplified by 
their scale-free nature. We follow the approach of R80 and K93, and 
attempt to reproduce the surface density on the unit circle, $\Sigma 
(R=1,\phi)$.  We restrict ourselves to the first quadrant and divide 
the unit circle into $N$ equally spaced azimuthal cells, with surface 
density $\Sigma_{\rm cell}(i)=\Sigma (R=1,\phi _i)$, $(i=1,..,N)$.  We 
define $\Sigma_{\rm orb}(i,j)$ as the mass contribution of the $j$th 
orbit ($j=1,..,M$) to the $i$th cell. The model is self-consistent if 
\begin{equation}
\sum _{j=1}^{M} \Sigma_{\rm orb}(i,j) w(j) = \Sigma_{\rm cell}(i),
	   \qquad i=1,\ldots,N,
\label{eq:self-consistent}
\end{equation} 
subject to the condition 
\begin{equation} 
w(j) \ge  0, \qquad j=1,..,M.
\label{eq:weights}
\end{equation}
The weights $w(j)$ are a numerical representation of the phase-space
distribution function that produces the surface density $\Sigma$ in
the potential $V$ (cf.\ Schwarzschild 1979; K93).

We take $M\gg N$, so that the system of equations 
(\ref{eq:self-consistent}) and (\ref{eq:weights}) is overdetermined. A
solution of (\ref{eq:self-consistent}) subject to the constraints 
(\ref{eq:weights}) can be found by standard linear programming (LP) 
methods such as the simplex method (Press et al.\ 1992) or the
non-negative least squares (NNLS) method (Pfenniger 1984). The NNLS 
method is useful when kinematic constraints are included 
(e.g., Rix et al.\ 1997). We are interested only in reconstructing 
the surface density distribution, and we employ the simplex method 
through minimizing the {\it non-self-consistency index} $Y$ defined 
in K93 as 
\begin{equation}  
Y=\frac{1}{N \overline \Sigma} \sum_{i=1}^{N}  
     \left | \Sigma_{\rm cell}(i) - \sum_{j=1}^{M}  
			       \Sigma_{\rm orb}(i,j) w(j)  \right |,  
\label{eq:non-consistency}  
\end{equation}   
subject to the constraints $w(j) \ge 0$. The quantity $Y$ must vanish 
for self-consistent discs. $\overline \Sigma$ is the mean surface 
density on the unit circle for $0 \le \phi \le \pi /2$.   
 
The calculation of $\Sigma_{\rm cell}(i)$ is straightforward through 
eq.~(\ref{eq:surf-density}).  However, care needs to be taken with the 
computation of the orbital densities $\Sigma_{\rm orb}(i,j)$. At some 
arbitrary time, stellar orbits do not necessarily cross the unit 
circle. But, some of their scaled, similar families do.  The amount of 
mass that a star moving in the scaled orbit deposits in the azimuthal 
bin $\Delta \phi$ (on the unit circle) is proportional to $\Delta\tau 
/\Delta\phi$. According to eq.~(\ref{eq:scaled-time}), this is equal 
to $R^{\frac {\alpha}2 -1} \Delta t/\Delta \phi$ where we have set 
$\tilde R=1$.  Thus, having calculated the orbital density at 
$[R(t),\phi (t)]$, one can readily find the amount of mass contributed 
to the $i$th cell of unit circle through multiplying the obtained 
density by $R^{\frac {\alpha}2-1}$. For $\alpha=0$, this factor is 
$R^{-1}$ as noted by R80 and K93.

\subsection{Curvature condition} 
\label{sec:curvature-flat} 
 
Consider an azimuthal element of length $\Delta \phi$ on the unit 
circle. The surface density per unit length of this element will be 
$S(\phi_i) \Delta \phi$ where $\phi_i$ is the angular position of the 
centroid of the $i$th element. The mass contribution of the $j$th 
orbit family to the assumed azimuthal cell is proportional to $(\Delta 
t/T)_j R_j^{\alpha /2 -1}$ where $(\Delta t /T)_j$ is
the fraction of the total integration time $T$ that a test star of
orbit family $j$ spends in the angular sector $(\phi _i,\Delta \phi)$.
The model is self-consistent when (cf.\ eq.~[\ref{eq:self-consistent}])
\begin{equation}
\sum _{j=1}^{M}w(j) \left \langle 
(\Delta t /T)_j R_j^{\alpha /2-1} \right \rangle =
S(\phi _i) \Delta \phi,~~~i=1,...,N,  
\label{eq:discrete-self-consistent}  
\end{equation}
where $M$ is the number of orbits, $N$ is the number of azimuthal 
cells on the unit circle, and the $w(j) \ge 0$ are weights.
We have taken time averages of orbital densities (denoted by
angles) for all the times when an orbit returns to the same 
sector. We can rewrite (\ref{eq:discrete-self-consistent}) 
in the form
\begin{equation}
\sum _{j=1}^{M}w(j) \left \langle 
\frac {(\Delta t /T)_j } {\Delta \phi}
\frac {R_j^{\alpha /2-1} } { S(\phi _i) }
\right \rangle =1, \qquad i=1,...,N.  
\label{eq:alt-self-consistent}  
\end{equation}
In the limit, as $\Delta \phi \rightarrow 0$ so that
$N \rightarrow \infty$ and we generally also need to take
$M \rightarrow \infty$, one obtains 
\begin{equation}  
\sum _{j=1}^{\infty} w(j) \left \langle
\frac {R_j ^{\alpha /2 -1} }
{ \dot \phi _j S(\phi _j) } \right \rangle = 1,
\qquad \forall \phi _j \in \left [0,\frac {\pi}2 \right ].
\label{eq:cont-limit-consistent}  
\end{equation}  
This equation is the continuous version of (\ref{eq:self-consistent})
in that the cells of configuration space are shrunk to zero size and
{\it instantaneous} orbital densities are taken into account.
The orbit families which occur in (\ref{eq:cont-limit-consistent})
can be regular or chaotic.

Eq.~(\ref{eq:cont-limit-consistent}) can be expressed in terms
of the angular momentum $J_j=R_j^2 \dot \phi _j$ as 
\begin{equation}  
\sum _{j=1}^{\infty}w(j) \langle \Gamma \rangle =1, \qquad  
\Gamma = \frac 1{\Vert J_j \Vert \mu (R_j,\phi _j)},  
\label{eq:def-gamma}  
\end{equation}  
where we have defined  
\begin{equation} 
\mu (R_j,\phi _j)= R_j^{-\gamma}S(\phi _j), \qquad  
\gamma =1+\frac {\alpha}{2}. 
\label{eq:rj-gamma} 
\end{equation} 
We assume $S(\phi)$ is smooth. Since $J_j$ is also smooth by the 
equations of motion (\ref{eq:eqs-motion}), we can follow ZCZ, and take 
the second derivative of (\ref{eq:def-gamma}) with respect to 
$\phi _j$. This gives 
\begin{equation}  
\sum _{j=1}^{\infty}w(j) \left \langle
\frac {{\rm d}^2 \Gamma}{{\rm d}\phi _j^2}
\right \rangle = 0,~~ \forall \phi _j \in
\left [0,\frac {\pi}{2}\right ].
\label{eq:second-derivative} 
\end{equation} 
This relation is a {\it necessary} condition for $\Gamma$ and will be 
satisfied if at least some orbit families have positive ${\rm d}^2 
\Gamma /{\rm d}\phi _j^2$ and others a negative value. If all ${\rm 
d}^2 \Gamma /{\rm d}\phi _j^2$ are either strictly positive or 
strictly negative, then the condition can only be satisfied by having 
positive as well as negative $w(j)$, so that the model is not 
self-consistent. 
 
As pointed out by ZCZ, the evaluation of ${\rm d}^2 \Gamma / {\rm 
d}\phi _j^2$ is simplified considerably near the major axis of the 
disc.  Hence, we investigate (\ref{eq:second-derivative}) at 
$\phi_j=0$. We drop the subscript $j$ of $\phi$ and $R$ but keep in 
mind that these variables are not the same for different orbits. 
Following ZCZ we obtain 
\begin{equation}  
\frac {{\rm d}^2\Gamma}{{\rm d}\phi ^2}=\frac {\Gamma}{K_{\phi}}  
\left [Q_V-(1+\gamma-\gamma \lambda) \right ],  
\label{eq:criterion-one}  
\end{equation}  
where  
\begin{equation}  
\lambda=\left[(1+\gamma)K_R+Q_{\mu}K_{\phi} \right ]_{\phi =0},
\label{eq:criterion-two} 
\end{equation}
and
\begin{equation}
K_R=\frac {\dot R ^2}{R \frac {\partial V}{\partial R}}, \qquad
K_{\phi}=\frac {(R \dot \phi)^2}{R \frac {\partial V}{\partial R}}.
\label{eq:criterion-three}
\end{equation}
The quantities $Q_V$ and $Q_\mu$ are defined as
\begin{equation}
Q_V=1+\frac {\frac {\partial^2 V}{\partial \phi^2} }
       {R \frac {\partial V}{\partial R}} \Big\vert_{\phi =0}, \qquad 
Q_\mu=1+\frac {\frac {\partial^2\mu}{\partial \phi^2} }
      {R \frac {\partial \mu}{\partial R}} \Big\vert_{\phi =0}.
\label{eq:criterion-four}
\end{equation} 
If ${\rm d}^2 \Gamma/{\rm d}\phi^2$ of eq.~(\ref{eq:criterion-one}) 
does not
have a zero, then the associated disc cannot be self-consistent.

The condition for self-consistency derived by ZCZ appears to be 
identical to eq.~(\ref{eq:criterion-one}), but their derivation did 
not take into account the scaling of the orbits, which caused them to 
use $\mu=\mu_{\rm ZCZ}=R^{-\gamma} S(\phi)$ and $\gamma=\gamma_{\rm 
ZCZ}=1-\alpha$. All the formulae given in ZCZ for the second 
derivative of $\Gamma$ are still applicable --- and are indeed 
identical to eqs~(\ref{eq:criterion-one}) to (\ref{eq:criterion-four}) 
--- but they need to be evaluated with $\mu$ and $\gamma$ given in 
eq.~(\ref{eq:rj-gamma}).  The definitions (and results!) agree only 
when $\alpha=0$, i.e., for discs with logarithmic potentials (i.e., 
flat rotation curves) such as the models studied by K93. 
 
\begin{figure*} 
\centerline{\hbox{\epsfxsize=8.4truecm\epsfbox{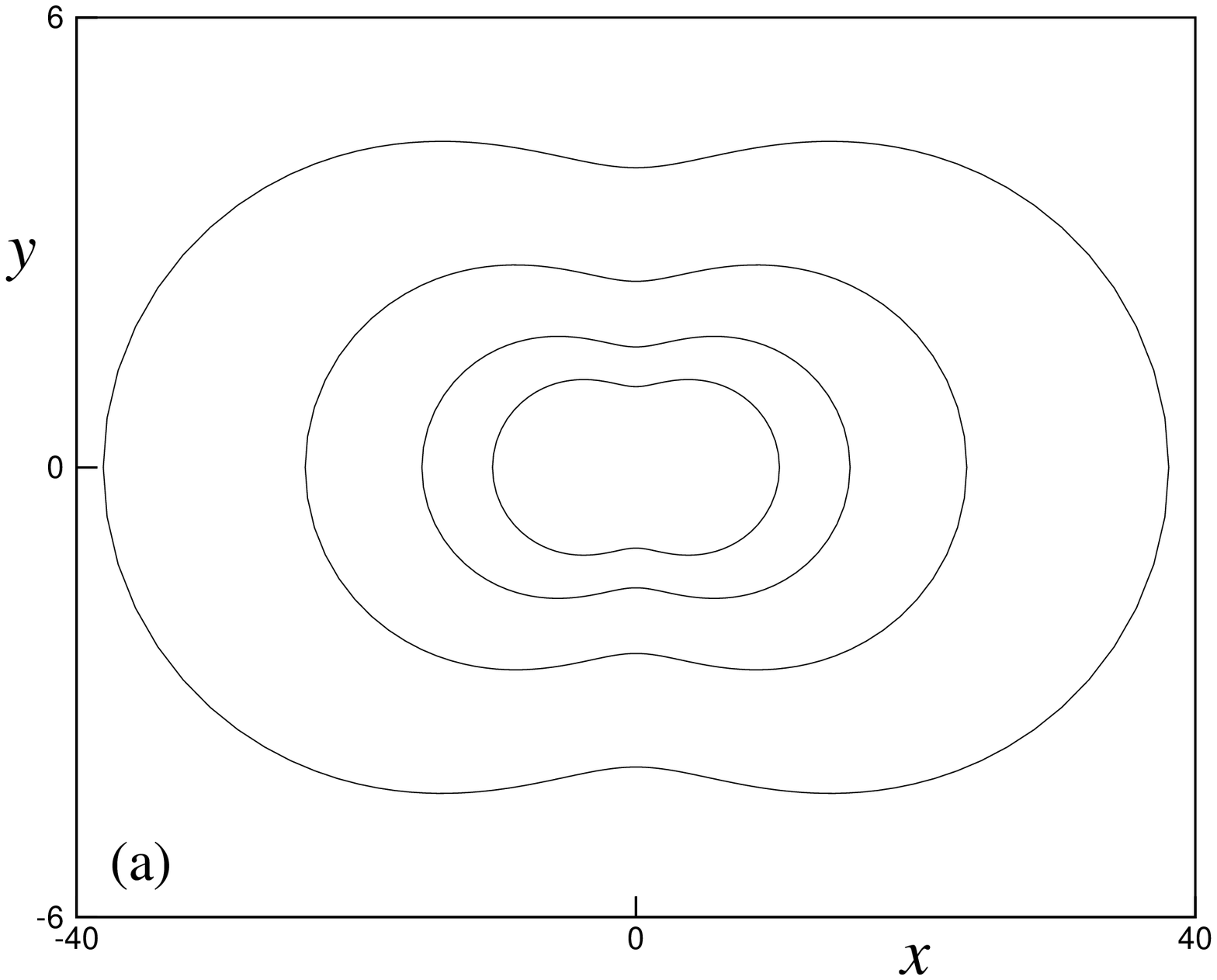}\hspace{0.2cm} 
\hfill
                  \epsfxsize=8.4truecm\epsfbox{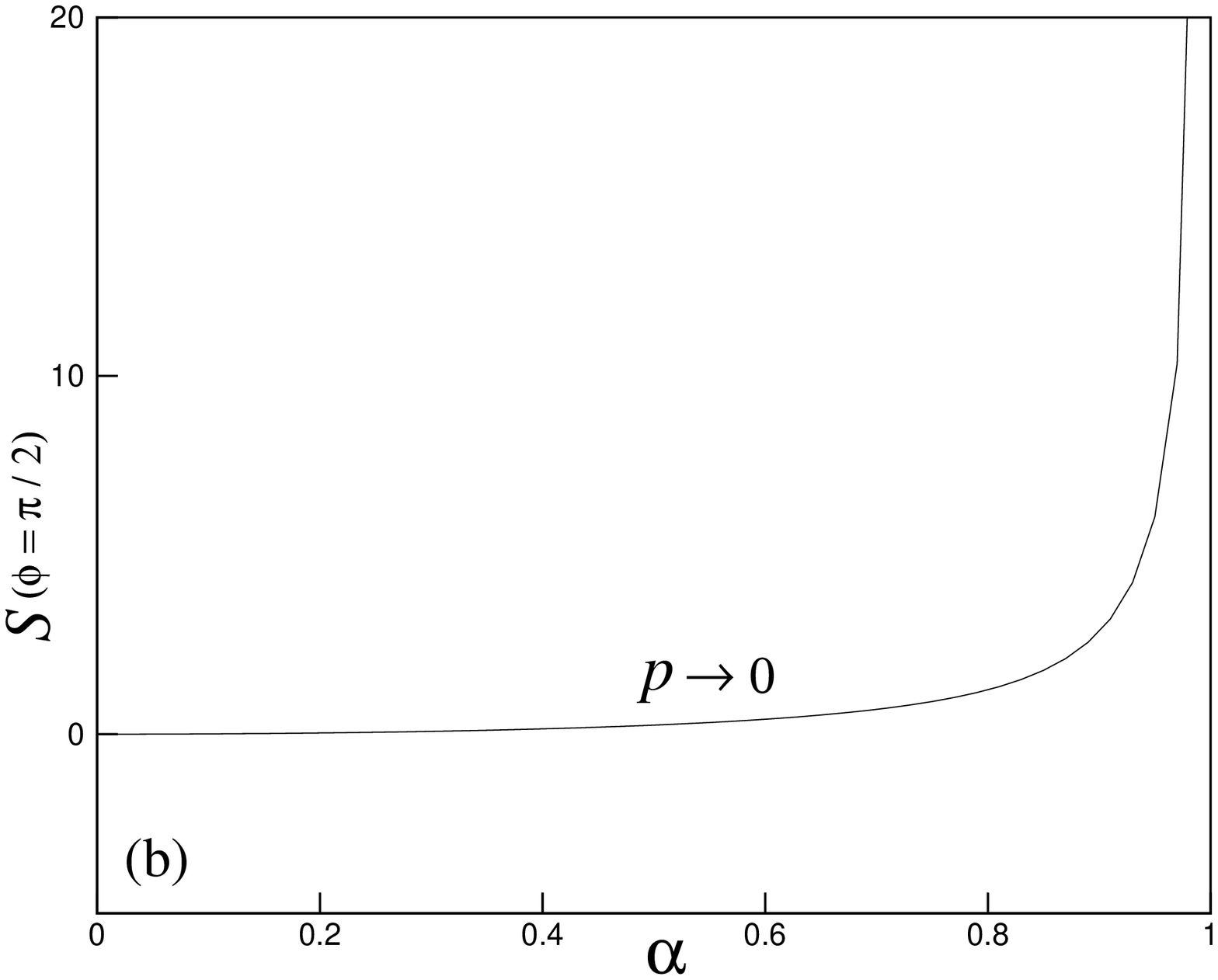}}} 
\caption{Panel {\em a} shows the surface density isocontours of the
power-law disc with $p=0.2$ and $\alpha=0.8$.  Panel {\em b} shows 
the lower bound of $\Sigma$ that occurs at $\phi =\pi/2$ as $p
\rightarrow 0$, as a function of $\alpha$.}
\end{figure*}

\section{The power-law discs} 
\label{sec:power-law-discs} 
 
In order to compare the results of the curvature condition with 
numerically constructed discs, we now investigate a set of elongated
discs with a very simple potential which makes it feasible to
construct a large number of numerical models.

\subsection{Potential-density pair} 
\label{sec:power-law-discs-props} 
 
We take a potential of the form (\ref{eq:potential}), with  
\begin{equation} 
P(\phi)= (p^2 \cos^2\phi + \sin^2\phi)^{\alpha \over 2}, 
\label{eq:ptheta-power-law} 
\end{equation} 
so that the equipotentials are similar concentric ellipses with axis
ratio $p$. The angular dependence $S(\phi)$ of the associated surface
density (\ref{eq:surf-density}) is derived as a series expansion in
Appendix A, and given in eqs~(\ref{eq::power-law-density}) and
(\ref{eq:explicit-surf-den}). Figure~1{\em a} shows the surface
density isocontours for $p=0.2$ and $\alpha=0.8$. They are slightly
dimpled near the minor axis ($\phi=\pi/2$), where $S(\phi)$ has a
minimum. The dimples become stronger as $p$ is decreased to zero.  We
have numerically evaluated $S(\pi/2)$ in this limit. Figure~1{\em b}
shows that the result is positive for all $\alpha$, and hence
demonstrates that the surface density is positive for all values of
$0\le p \le 1$ and $0 \le \alpha  \le 1$.
\begin{figure*}  
\centerline{\hbox{\epsfxsize=8.5truecm\epsfbox{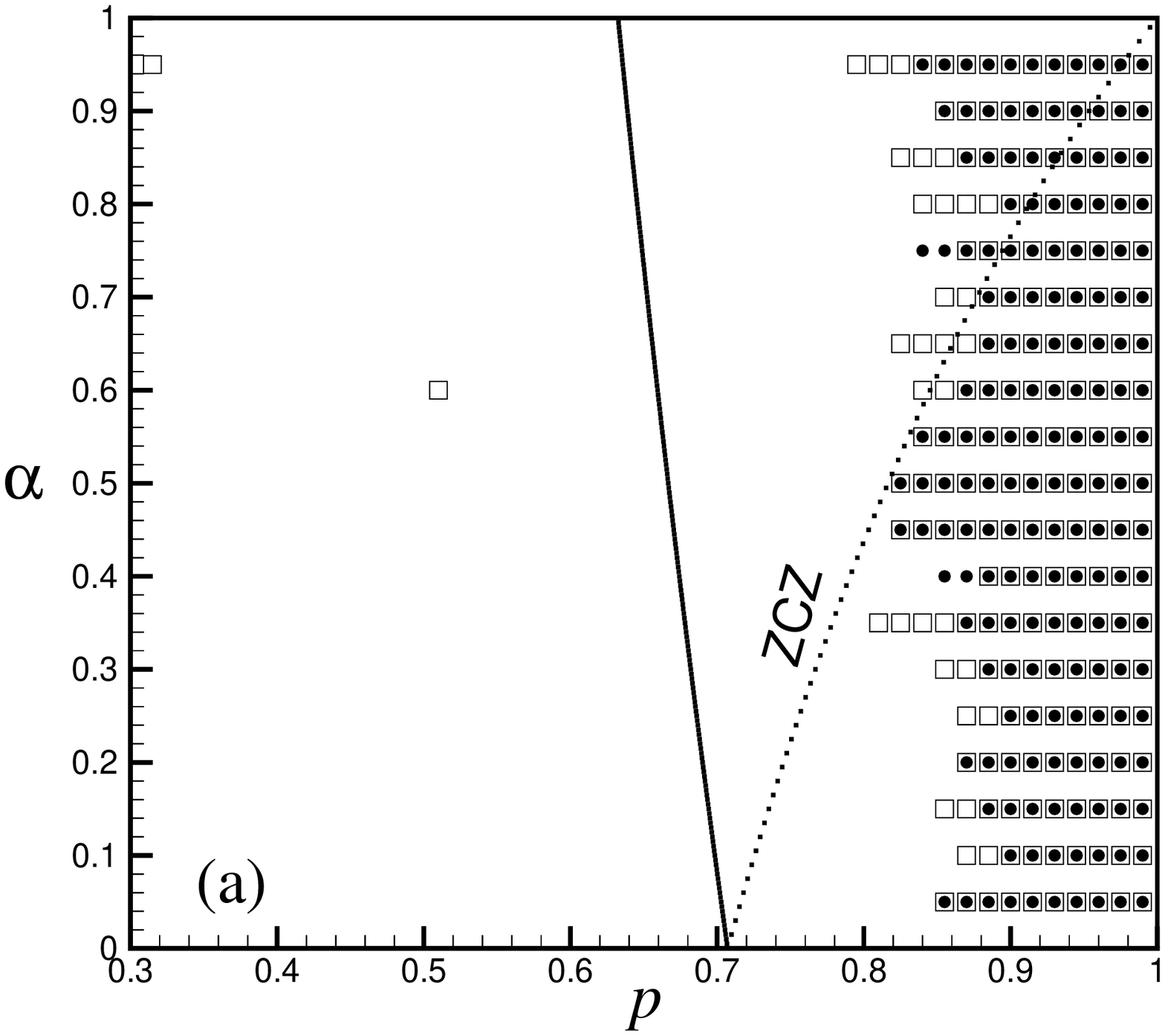}\hspace{0.3truecm} 
		  \epsfxsize=8.5truecm\epsfbox{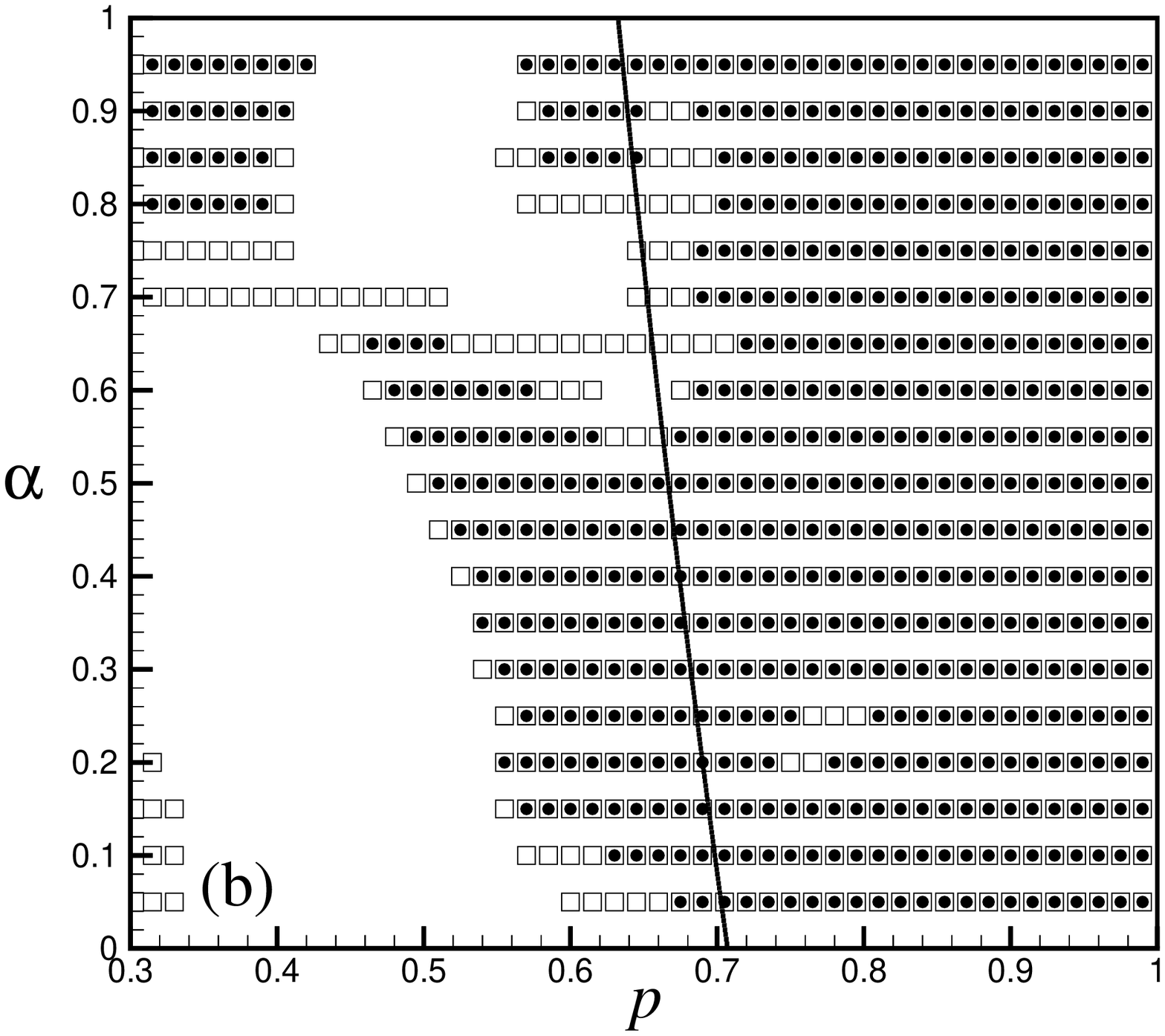}}}
\caption{The $(p,\alpha)$ parameter space of power-law discs. For the
marked points the reconstruction of the model density by
Schwazrschild's numerical orbit superposition method was successful by
assuming the convergence condition $\tilde Y<\sigma_N$, defined in
equation (\ref{eq::standard-deviation}) (a) Squares correspond to 
$N=M_2=45$ and filled circles correspond to $N=M_2=90$ (b) Squares and 
filled circles correspond to the pairs $(N=45,M_2=180)$ and 
$(N=140,M_2=560)$, respectively. In all cases we have set $M_1=80$. 
The solid curve is the limit (\ref{eq:condition-power-law}) of the 
region of non-existent models provided by the curvature condition 
Non-consistent models are to the left of the solid curve. 
The dotted line indicates the limit derived by application of the 
(erroneous) condition from ZCZ. The distribution of points in Panel
{\em b} is discussed in \S3.4}
\end{figure*}

\subsection{The curvature constraint} 
\label{sec:curvature-condition-power-law}
 
The series expansion
(\ref{eq::power-law-density}, \ref{eq:explicit-surf-den})
for $S(\phi)$ can be used to show that the factor $Q_\mu$ in
eq.~(\ref{eq:criterion-two}) is non-negative for
$0 \leq p \leq 1$ and $0 \leq \alpha \leq 1$. As a result,
$0<\lambda _{\rm min}\le \lambda \le \lambda _{\rm max}<\infty$.
Since $\gamma=1+\alpha/2> 0$, we find that the second derivative
of $\Gamma$ will change sign when
\begin{equation}  
1+\gamma-\gamma \lambda _{\rm max} < Q_V <
1+\gamma-\gamma \lambda _{\rm min}.  
\label{eq:criterion-power-law}  
\end{equation}  
This is a necessary but not sufficient condition for self-consistency. 
This curvature condition imposes a constraint on the curvatures of $V$ 
and $\mu$ on the major axis. For $\lambda _{\rm min} =0$, a 
non-trivial limit of (\ref{eq:criterion-power-law}) is obtained as 
\begin{equation}  
p > \frac 1{\sqrt{2+\frac {\alpha}2}},  
\label{eq:condition-power-law}  
\end{equation}  
which defines the boundary curve between the allowed and forbidden
zones of the parameter space (solid curve in Figure~2{\em a}). 
Specifically,
$p > 0.707$ for $\alpha=0$ and $p > 0.633$ when $\alpha=1$. This
corresponds to minimum axis ratios of the contours of constant surface
density of 0.500 and 0.507, respectively.

\subsection{Numerical models} 
\label{sec:schwarzschild-models} 
 
We now compare the results derived from the curvature condition with
numerical models, constructed with Schwarzschild's method as outlined
in \S\ref{sec:schwarzschild}. We divided the parameter space into 48
cells in the $p$-direction (from $p=0.3$ to $p=1$) and 20 cells in the
$\alpha$-direction (from $\alpha=0$ to $\alpha=1$). For each model we
generated a library of $M$ orbits containing $M_1$ loops (flat tubes)
and $M_2$ orbits with zero initial velocities (along with their
reflections with respect to the coordinate axes) dropped at $M_2$
equally spaced azimuths on the unit circle between $\phi=0$ and
$\phi=\pi/2$ (these include fishes, bananas, pretzels, $\ldots$, cf.\
Miralda-Escud\'e \& Schwarzschild 1989). We compute the surface 
density of the $i$th cell at $\phi _i=\frac {\pi}{2N}(i-\frac 12)$ 
($1\le i \le N$), and uniformly distribute the initial positions 
of orbits (with zero initial velocities) on the unit circle. i.e., 
the $j$th orbit is dropped at $\phi _j=j\pi/[2(M_2+1)]$ where 
$1\le j \le M_2$. In this way, initial positions of orbits will not 
coincide with the boundaries of cells. Our loop orbit library consists 
of both thin and thick orbits, which are launched on the major axis 
with $y(0)=\dot x(0)=0$, $x(0)=1$ and $\dot y(0)\ge \dot y_{\rm min}>0$.
For $\dot y(0)<\dot y_{\rm min}$, the launched orbit will not belong 
to the 1:1 resonant island. We find $\dot y_{\rm min}$ numerically.

In order to calculate the orbital densities, we integrated the
equations of motion for up to $T=120$ galactic years using the RK78
routine of Fehlberg (1968), which guarantees the preservation of
energy with an accuracy of $10^{-12}$. For a regular orbit, $T$ should
ideally be the time by which the orbit becomes dense on its invariant
torus, but this may require very long integrations for near-resonant
and chaotic orbits. We therefore decided to follow Cretton et al.\
(1999), and to fix $T$ for all orbits to a value that is sufficiently
large for the remaining variations in the orbital density not to
influence the main properties of the dynamical model. Experiments
showed that for $T \ge 120$ the remaining `noise' in the modeling
procedure is dominated by the representation of phase space through a
discrete grid, and not by the properties of the individual orbit
densities. We therefore chose $T=120$ throughout.

\begin{figure}  
\centerline{\hbox{\epsfxsize=8truecm\epsfbox{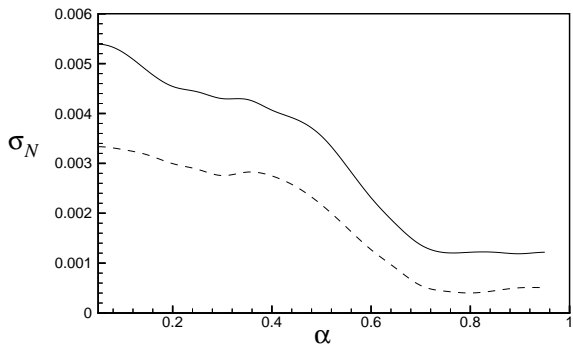}}}
\caption{Variation of $\sigma _N$ versus $\alpha$.
The solid and dashed curves correspond to $M_2=N=90$
and $M_2=4N=560$, respectively. For each $\alpha$,
models having $\tilde Y<\sigma _N$ are accepted to be 
self-consistent.}
\end{figure}

We ran our LP code for different choices of the number $N$ of angular
cells, considered two values of the ratio $M_2/N$, and took $M_1=80$
loops in all cases. The results are illustrated in Figure~2, which
shows all the points in the $(p, \alpha)$-plane for which the
numerical reconstruction of the model density was successful.
For self-consistent discs $Y$ should ideally become zero to
within machine precision. However, our numerical computations
show that for large values of $p$, the noise in the value of
$Y$ can have almost the same order-of-magnitude as $Y$ itself.
Thus, one needs a non-zero error threshold to pick up
self-consistent models. K93 defined a threshold value for
non-self-consistency based on the isotropic distribution function
of axisymmetric models, even though numerical experiments showed
that $Y$ is smaller for shallow axisymmetric cusps than for steep
cusps. We therefore adopted the following scheme to find a credible
threshold. For each value of $\alpha$, we determine the function
$Y(p)$. We then filter $Y(p)$ using the Savitzky-Golay smoothing
filter (Press et al.\ 1992) to obtain the new distribution
$\tilde Y(p)$ from which one can calculate the noise distribution
$N_Y(p)=Y(p)-\tilde Y(p)$.  The models that we consider viable have
$\tilde Y(p)<\sigma _N$ where $\sigma_N$ is the standard deviation
of the noise distribution. It is defined by
\begin{equation}
\label{eq::standard-deviation}
\sigma _N=\sqrt{\langle N_Y^2 \rangle - \langle N_Y \rangle ^2}.
\end{equation}
Figure 3 shows the variation of $\sigma_N$ versus $\alpha$
for $M_2=N=90$ and $M_2=4N=560$. According to this figure, shallow
cusps (with $\sigma_N\approx 0.001$) have more accurate Schwarzschild
models than steep ones (with $\sigma _N\approx 0.004$). We note that
most of our self-consistent models have $\tilde Y(p)\approx 10^{-8}$,
which is well within machine precision. For all successful models,
noise fluctuations have been less than 1\% of the global maximum of
$\tilde Y$. This indicates the acceptable quality of our modeling. 

Figure 2{\em a} summarizes the results of our model computations for
the case $M_2=N$, so that there is one zero-angular-momentum orbit
dropped in each cell. We started our simulations with $N=45$ (squares)
and increased $N$ until the set of self-consistent models (in the
parameter space) converged (filled circles; $N=90$). The lower bound
of the allowed range of $p$ increases as $\alpha$ tends to
zero. Figure 2{\em b} shows the same plane, but now for the case
$M_2=4N$. It is evident that orbits with zero initial velocities play
an important role in the numerical reconstruction of the model
density: by increasing $M_2$ relative to $N$, the zone of admissible
parameters grows considerably, a property that was also noted by
K93. We discuss the implications of this numerical result, and the
difference between panels {\em a} and {\em b}, in \S3.4.

In all models $\approx$90\% of the loop orbits have zero weight, which
indicates that they are less useful in the construction of
self-consistent models. This is as expected (e.g., K93) because the
loop orbits of cuspy systems are universally anti-aligned with the
potential and density isocontours, and they can hence only contribute
part of the density (cf.\ de Zeeuw, Hunter \& Schwarzschild 1987). As
their individual orbital densities are smooth, a modest number of
loops is sufficient to provide an accurate reconstruction of this part
of the model density. Accordingly, the results do not depend
significantly on the number $M_1$ of loops in the orbit catalog, and
our choice $M_1=80$ is adequate for all cases.

\subsection{Implications for Schwarzschild's method}
\label{sec:schwarzschild-implications}

Figure~2{\em a} shows that the possible region for self-consistent
models predicted by the curvature condition agrees well with the
results of the numerically constructed models when $M_2 = N$: all of
the numerically feasible models lie (well) inside the possible zone
when we use $M_2=N$. By contrast, when $M_2=4N$, many of the
numerically acceptable solutions lie in the region forbidden by the
curvature condition, even though the value of $\sigma _N$ is smaller,
so the condition on $Y$ is more stringent.

The choice $M_2>N$ is not consistent with the continuum limit
expressed by eq.~(\ref{eq:cont-limit-consistent}); as an azimuthal
cell is shrunk to zero size, it becomes theoretically impossible to
drop more than one orbit with zero initial velocity from that cell (it
is still possible to launch an arbitrary number of orbits with
non-zero initial velocities from the same point). In the numerical
simulations with $M_2>N$, more orbital `corners' are available per
cell, which brings a degree of flexibility to the LP code that in turn
results in more solutions that appear acceptable. Many of these are
clearly ruled out by the analytic curvature condition. 

The requirement $M_2=N$ also arises naturally in separable models,
where the integral equation for the weights of the box orbits is
solved analytically by assigning one box-orbit corner to each cell (de
Zeeuw, Hunter \& Schwarzschild 1987; Statler 1987). While box orbits
have one corner in the first quadrant, regular boxlets may have more,
so that the rule $M_2=N$ may count some orbits more than once (K93).
We conclude that in applications of Schwarzschild's method care should
be taken that the number of orbits used is matched properly to the
grid of cells on which the density is reproduced.

\begin{figure*}  
\label{fig:poincare} 
\centerline{\hbox{\epsfxsize=3in\epsfbox{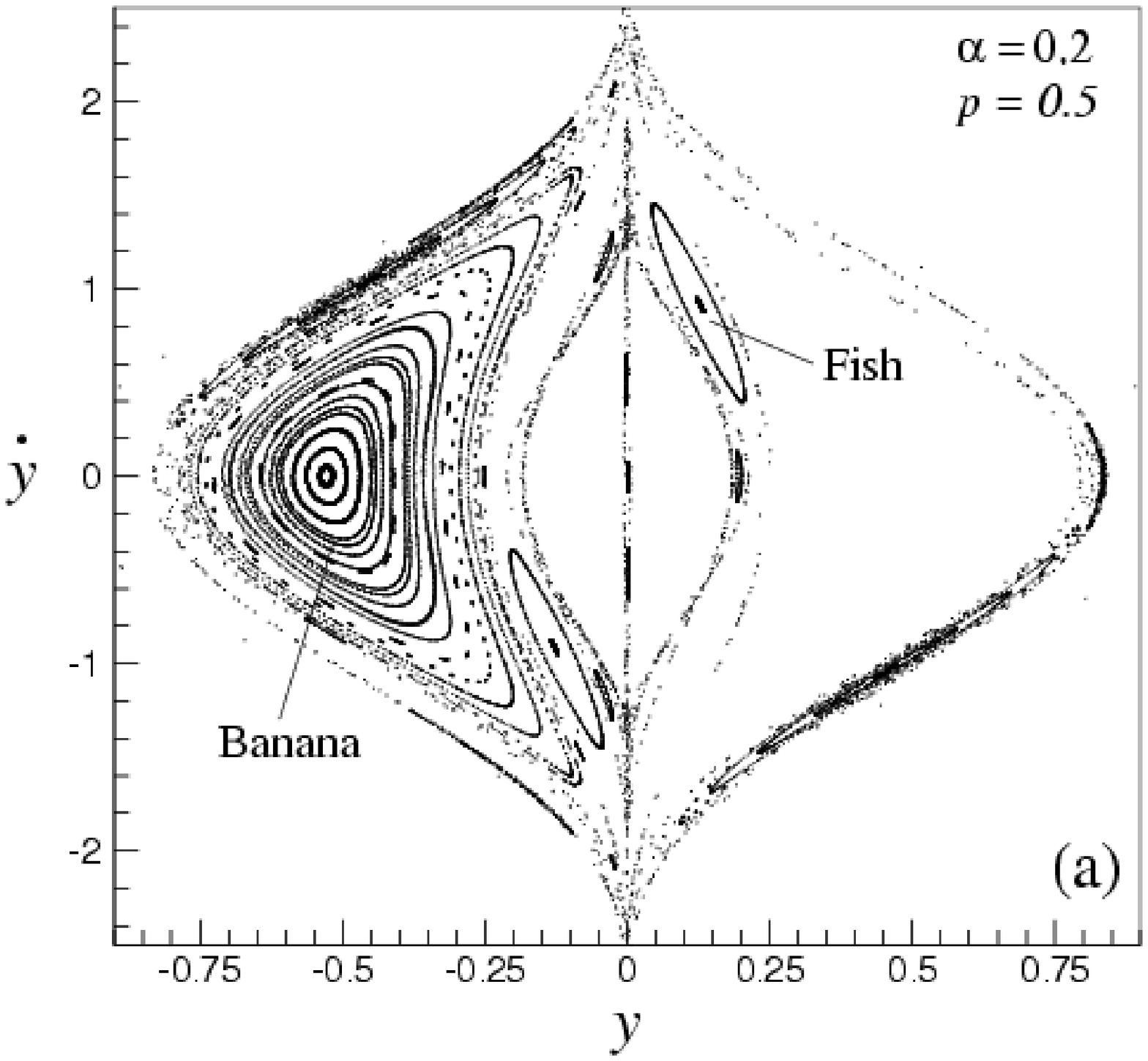}\hspace{0.6truecm} 
                  \epsfxsize=3in\epsfbox{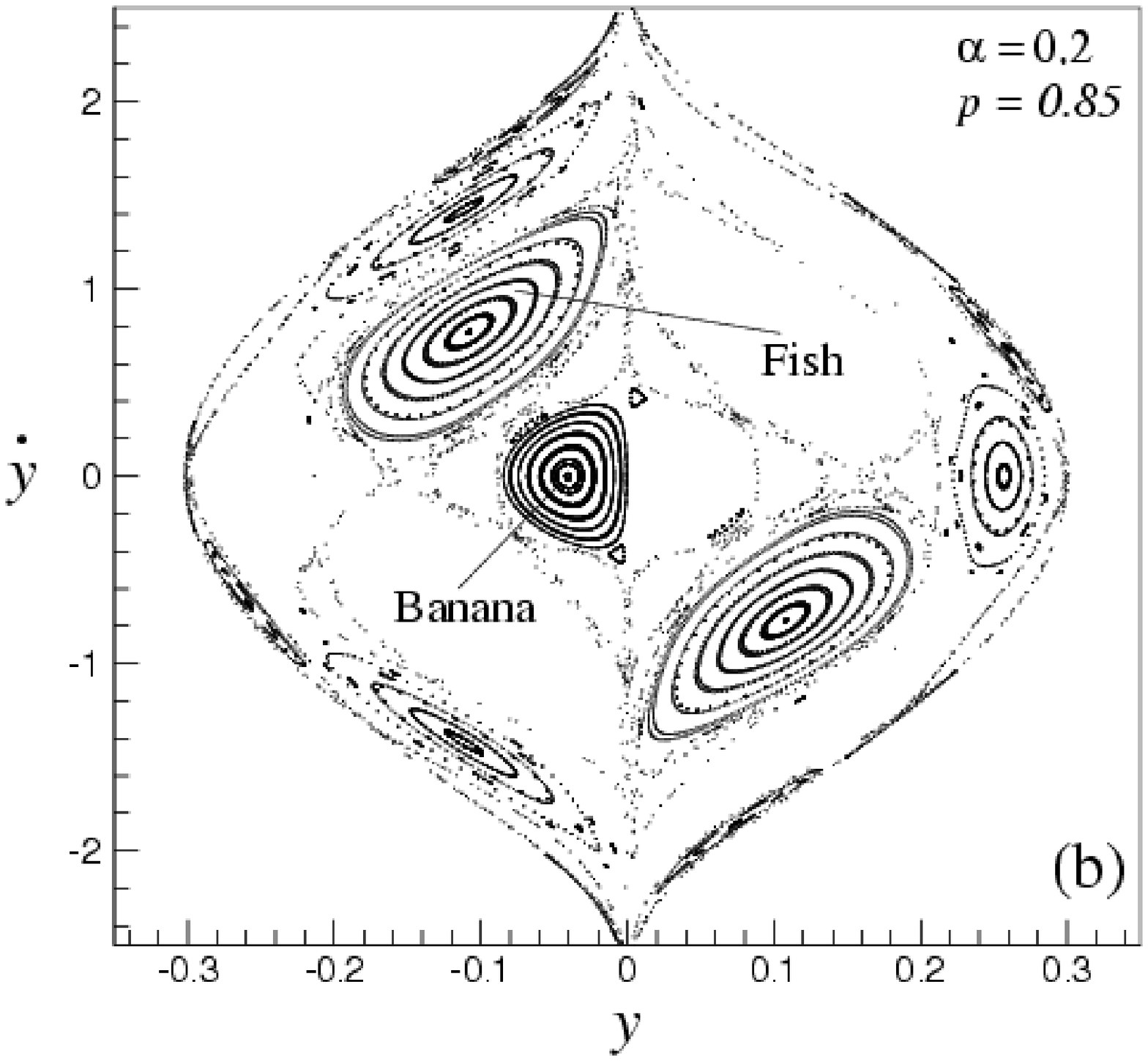}}} 
\bigskip 
\centerline{\hbox{\epsfxsize=3in\epsfbox{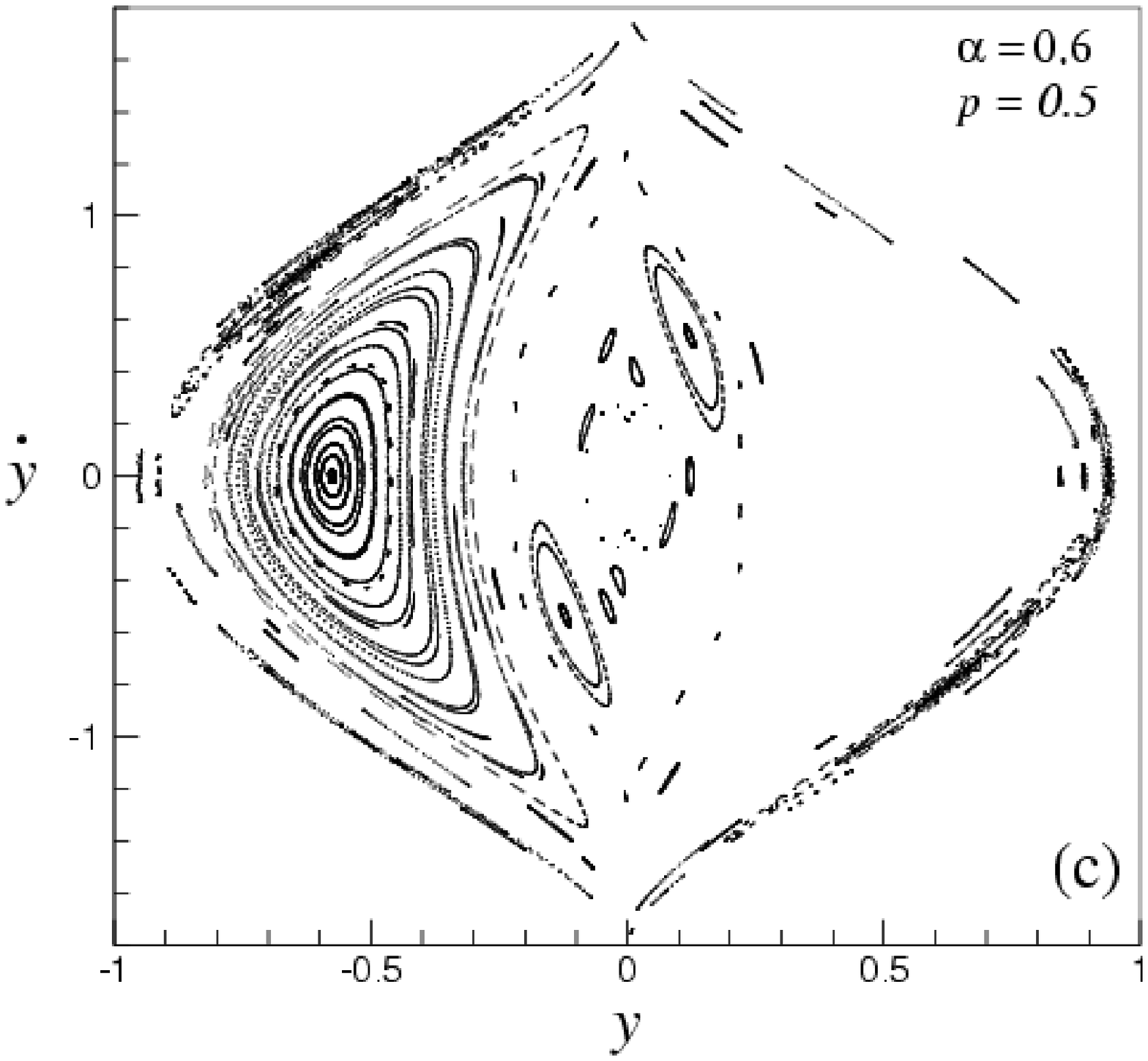}\hspace{0.6truecm} 
                  \epsfxsize=3in\epsfbox{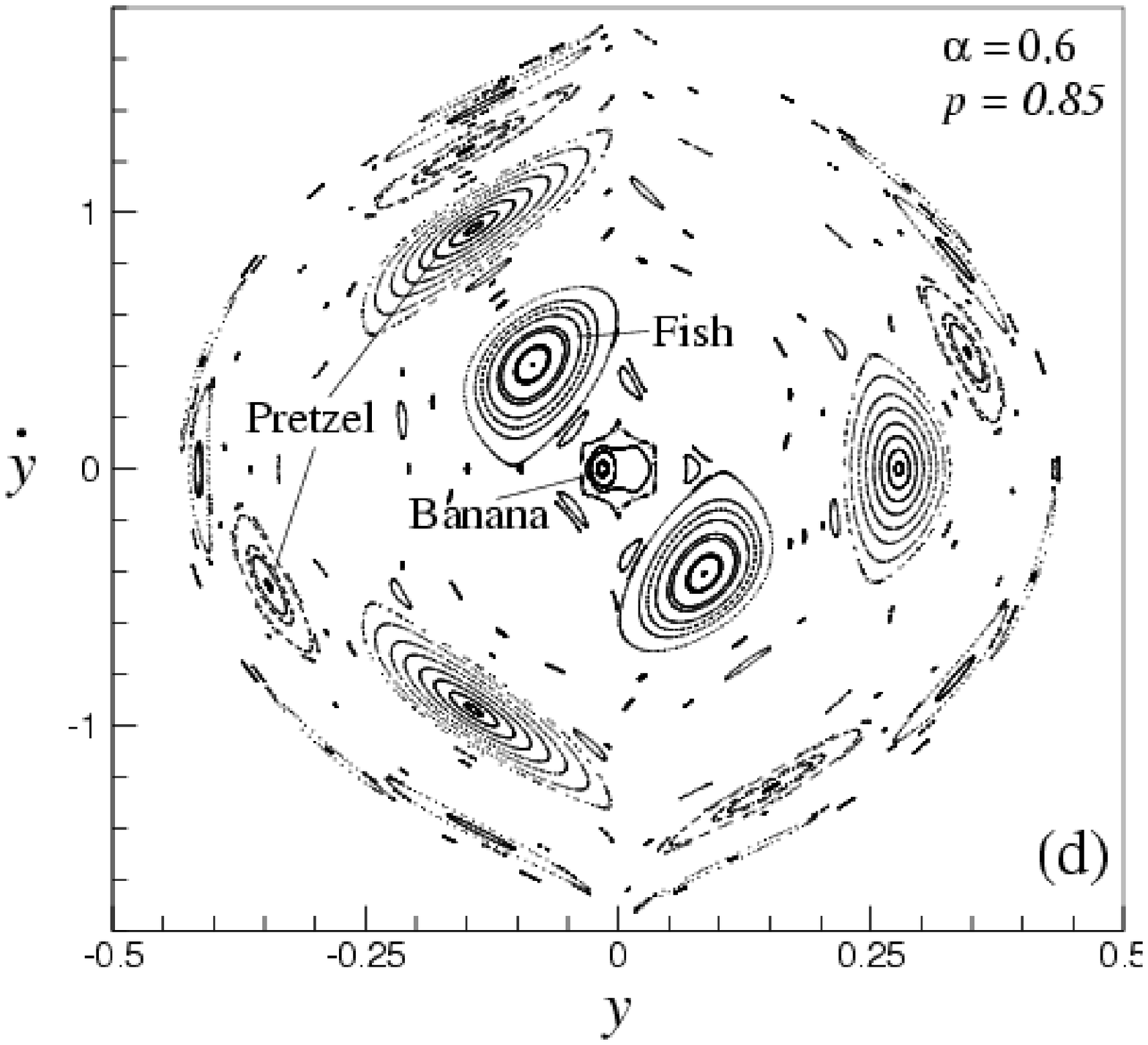}}} 
\bigskip 
\centerline{\hbox{\epsfxsize=3in\epsfbox{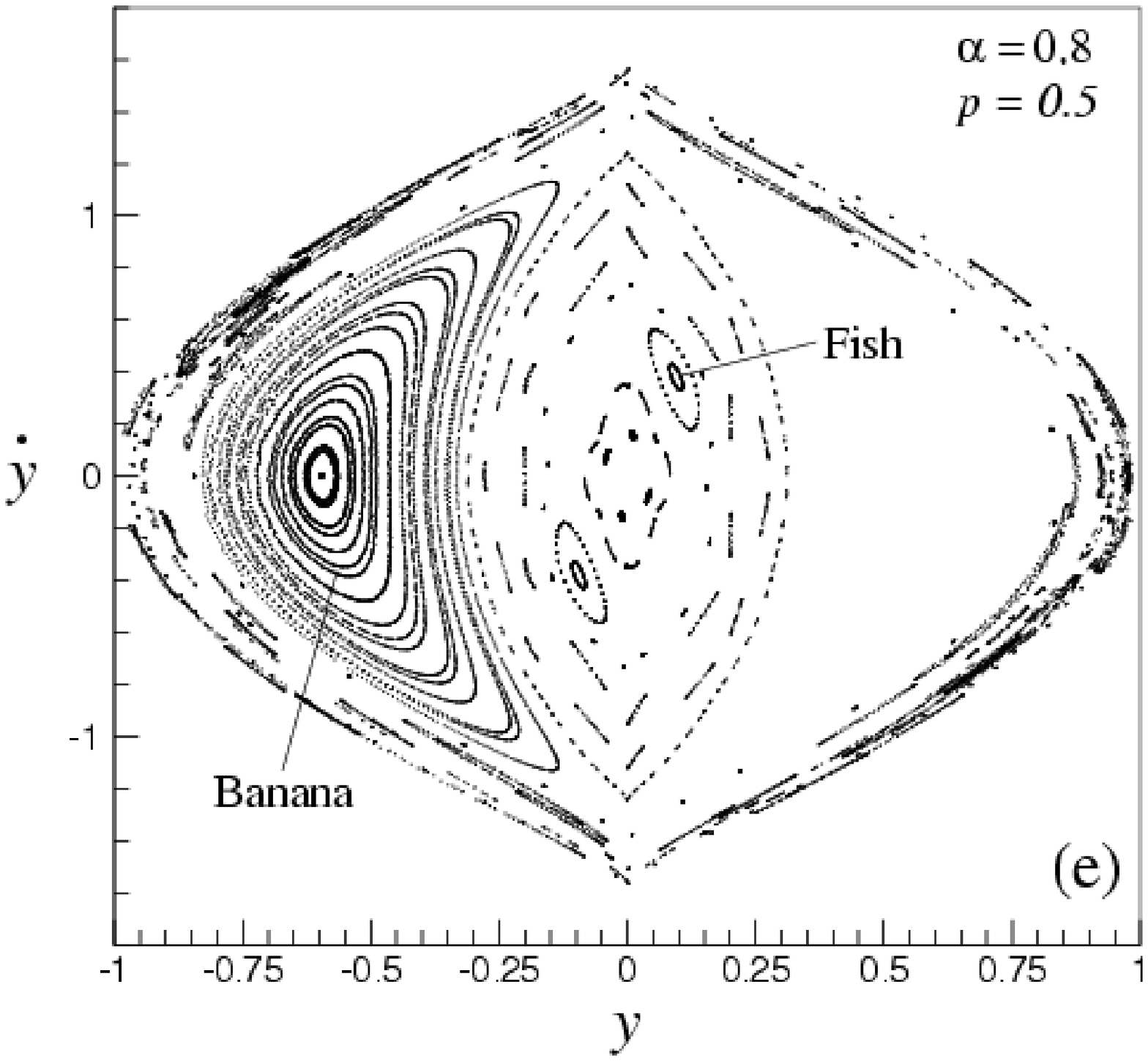}\hspace{0.6truecm} 
                  \epsfxsize=3in\epsfbox{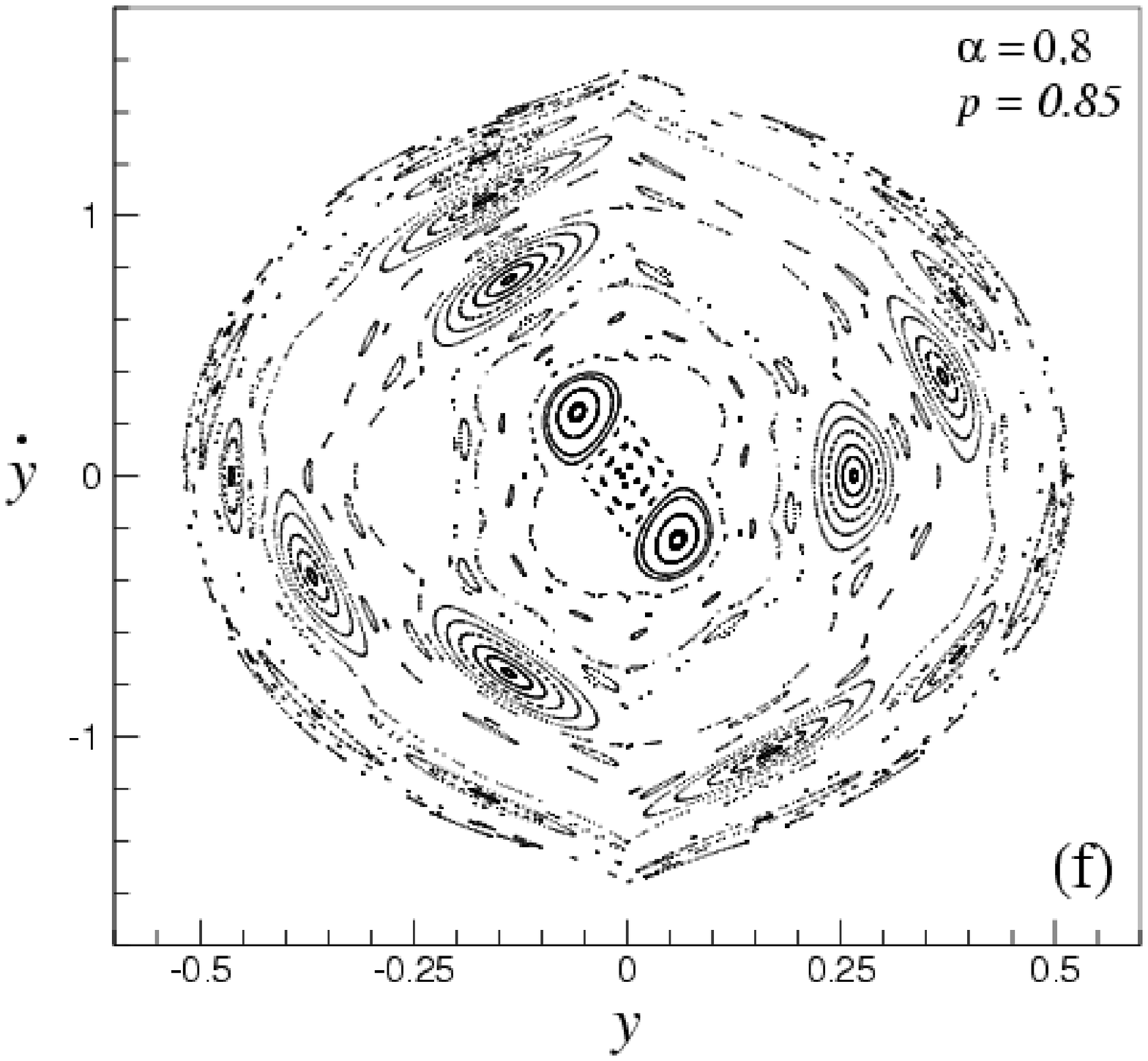} }} 
\caption{Poincar\'e maps of the orbits with zero initial velocities in 
the power-law discs. Orbits are dropped from equally-spaced azimuths 
between $\phi=0$ and $\phi=\pi/2$. In all cases we have generated the 
surfaces of section for $M_2=90$ orbits and have set the orbital 
energy $E=\frac 12 v^2 + V(R,\phi)$ equal to unity.} 
\end{figure*}
 
\subsection{Orbital structure} 
\label{sec:orbit-structure} 
 
To gain a better understanding of the origin of non-self-consistency,
we constructed the Poincar\'e maps of the orbits with zero initial
velocities for several choices of $\alpha$ and $p$. We integrated the
equations of motion in Cartesian coordinates, and sampled the phase
space variables every time an orbit crosses the $y$-axis with $\dot
x>0$. We have used the same library of orbits as in
\S\ref{sec:schwarzschild-models}, and dropped $M_2=90$ orbits from
equally spaced azimuths between $\phi=0$ and $\phi=\pi/2$. The energy
$E=\frac 12 v^2+V(R,\phi)$ of all orbits was set to $E=1$ and scaled
orbits were used if needed, because orbits dropped from the unit
circle do not all have the same energy. For each $\alpha$, we
generated the surfaces of section for two values of $p$: one in the
forbidden zone and the other in the allowed zone of the parameter
space explored by the curvature condition. The results are displayed
in Figure~4. The non-self-consistent models (with $p=0.5$) share an
interesting feature: the bulk of phase space is occupied by banana
orbits. By contrast, in models that are not ruled out by the curvature
condition, the phase space is dominated by fishes and high-resonant
orbits, while banana orbits are almost absent. This suggests that the
non-self-consistency is related to the existence of centrophobic
banana orbits that deposit much mass away from the major axis (as
discussed also by, e.g., Pfenniger \& de Zeeuw 1989; SZ98; ZCZ; Zhao
1999).

\begin{figure}
\centerline{\hbox{\epsfxsize=3.1in\epsfbox{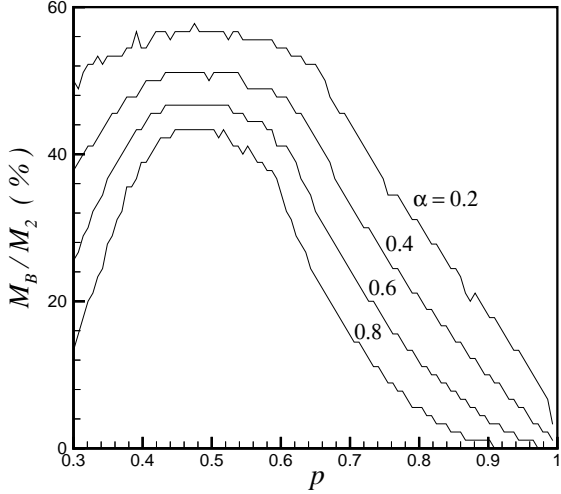}}}  
\caption{The variation of the relative population of banana orbits 
($M_{B/2}$) versus $p$ for several choices of $\alpha$. The global 
maximum of $M_{B/2}$ emerges in the neighborhood of $p=0.5$ for all 
values of $\alpha$.} 
\end{figure}

To pursue this idea further, we computed the number $M_B$ of banana
orbits, using the information in the Poincar\'e maps (in the $y$-$\dot
y$ cross section; banana orbits dropped from the first quadrant never
take $y>0$). Figure 5 shows the ratio $M_{B/2}=M_B/M_2$ (in
percentage) versus $p$ for several choices of $\alpha$.  $M_{B/2}$
decreases when $p \rightarrow 1$ and has a maximum near $p=0.5$. This
result is not unexpected, for banana orbits emerge when the
frequencies of the $x$- and $y$-oscillations are commensurate in the
ratio $\omega _x/\omega _y \approx 1/2$. To first order, $\omega
_x/\omega _y$ equals the axial ratio $p$ of the equipotential curves.
 
Figure~6 shows the computed values of the non-self-consistency index
$Y$ (for $N=M_2=90$ and $M_1=80$) versus $p$ for the models with
$\alpha=0.8$. The best-fitted polynomial curve has a similar variation
as the corresponding $M_{B/2}$ curve in Figure 5.  The models with
`banana-rich' orbital structures (corresponding to $p\approx 0.5$) are
{\it maximally} non-self-consistent. As $p\rightarrow 1$, $Y$
decreases like $M_{B/2}$ and converges to the required accuracy of
self-consistent models.\looseness=-2

According to Figure~4, only very thin layers of chaotic orbits occur
in the vicinity of hyperbolic fixed points (as also found by K93 for
the logarithmic discs). Therefore, they do not seem to play an
important role in the construction of the model density. This
conclusion is supported by the following arguments. The orbits of a
thin chaotic layer float around various resonant islands while
they remain `very close' to the outermost regular orbits (slow orbits)
of these islands. It is possible to approximately generate such
irregular orbits by a {\it random superposition} of their nearby slow
orbits. Thus, in the limit when a chaotic orbit is fully mixed in the
phase space, its orbital density approximately becomes equal to the
density of the `ensemble' of slow orbits. This means that a chaotic 
orbit enters Schwarzschild's method by adding (subtracting) a
constant weight to (from) the weights of slow orbits. Our numerical
computations confirm this point and reveal that our Schwarzschild
models do not depend on the integration time of `thin' chaotic
orbits. The above reasoning is no longer valid for `thick' chaotic
layers, which, however, do not emerge in our models. We conclude
that fishes and high-resonant orbits are the main building blocks
of the self-consistent discs.

\begin{figure}  
\centerline{\hbox{\epsfxsize=3.1in\epsfbox{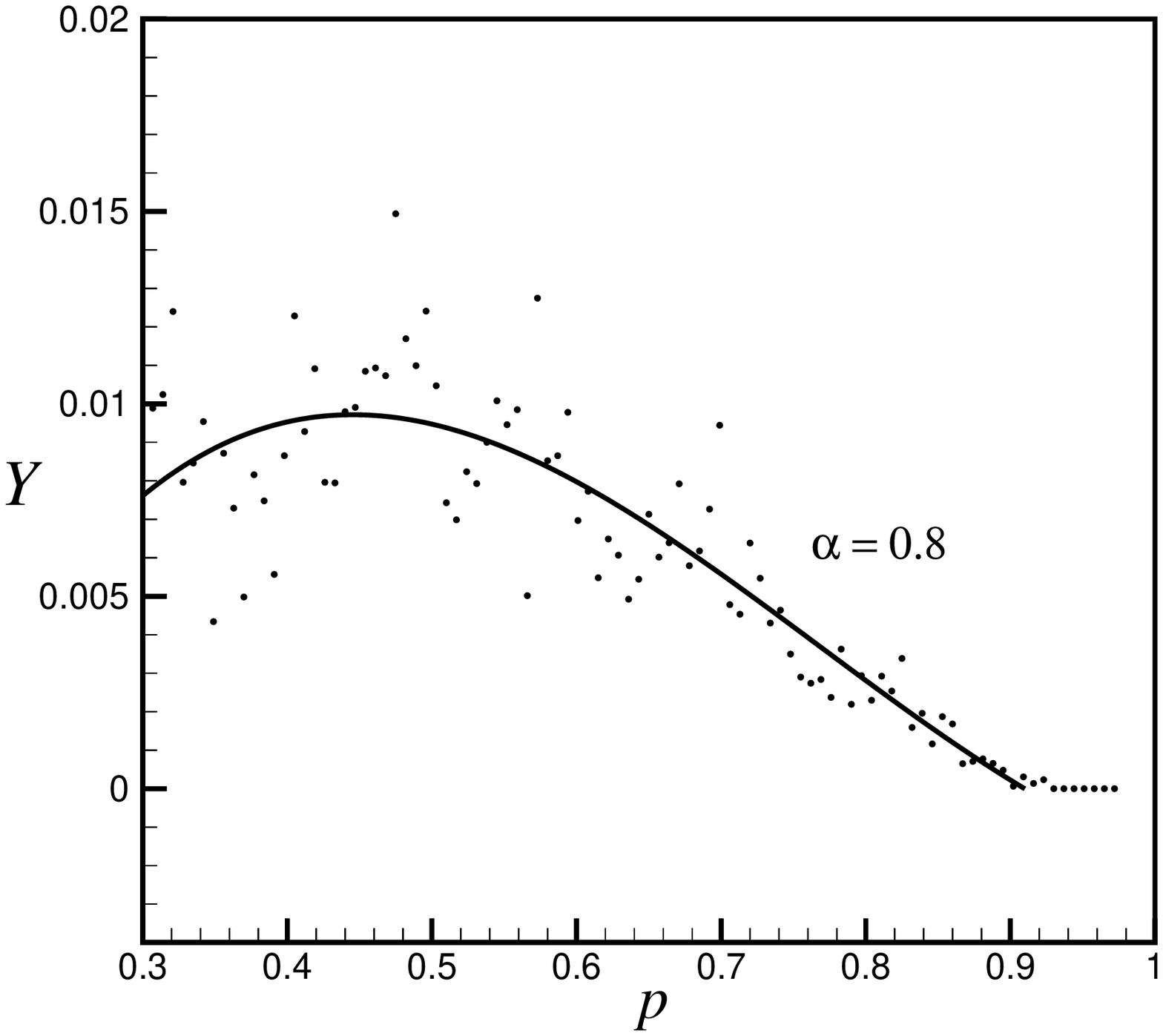}}} 
\caption{The variation of $Y$ versus $p$ for $\alpha =0.8$, $M_1=80$
and $N=M_2=90$. The solid line is the best fitted polynomial curve 
to the discrete numerical data shown by dots. Similar patterns occur
for the other values of $\alpha$. The non-self-consistency index
$Y$ has converged to zero with an accuracy better than
$\sigma _N\approx 0.001$ (for $p \gta 0.9$ we arrived at
$\tilde Y\approx 10^{-8}$).}
\end{figure}
 
\section{Other elongated discs}
\label{sec:other-discs} 
 
\subsection{Elliptic discs} 
\label{sec:elliptic-discs-props} 
 
We now consider discs with elliptic surface density distributions  
of the form (\ref{eq:surf-density}), with  
\begin{equation}  
S(\phi)  =(p^2\cos^2\phi +\sin^2\phi)^{\frac{\alpha -1}2}.  
\label{eq:elliptic-discs-surf}  
\end{equation}  
The gravitational potential in the plane of the disc is of the form 
(\ref{eq:potential}) with (e.g., Evans \& de Zeeuw 1992 eq.\ 5.2) 
\begin{equation}  
P(\phi) = \int\limits_{0}^{\infty}  
\frac {(p^2 \cos^2\phi \!+\! \sin^2\phi \!+\!u)^{\frac{\alpha}{2}}}
  {[(1\!+\!u)(p^2\!+\!u)]^{\frac{1+\alpha}{2}} 
                         u^{\frac{1}{2}}} \,  \, {\rm d}u,
\label{eq:elliptic-discs-pot}
\end{equation}
where $\Sigma_0$ and $V_0$ are related by $V_0 = 2\pi G \Sigma _0
p^\alpha /\alpha {\rm B} (\frac 12, \frac {1-\alpha}2)$ for 
$\alpha\not=0$. When $\alpha=0$ we have $V_0=2G\Sigma_0 R_F(1,p^2,0)$,
with $R_F$ the Carlson elliptic integral of the first kind (e.g.,
Press et al.\ 1992), and 
\begin{equation}
P(\phi) ={G\Sigma_0 \over V_0}
\! \int\limits_0^\infty 
  {\ln(p^2\cos^2\phi \!+\! \sin^2\phi \!+\! u) \over
   [(1\!+\!u)(p^2\!+\!u)u]^{\frac{1}{2}}} \, du. 
\label{eq:elliptic-discs-pot-spec} 
\end{equation}
The integrals (\ref{eq:elliptic-discs-pot}) and
(\ref{eq:elliptic-discs-pot-spec}) for the $\phi$-dependence of the
potential need to be evaluated numerically.

In order to apply the curvature condition, we evaluate $Q_\mu$ and
$Q_V$, and obtain
\begin{eqnarray}
Q_{\mu}  \!\!\! &=& \!\!\! \frac {3p^2\alpha +2(1\!-\!\alpha)} 
			     {(2+\alpha)p^2}>0, \nonumber \\   
Q_V \!\!\! &=& \!\!\!1+(1\!-\!p^2)\frac {P_2(p,\alpha)}{P_1(p,\alpha)},
\label{eq:elliptic-discs-qs}
\end{eqnarray}
where we have defined
\begin{equation}
P_i (p,\alpha) \!=\!\! \int\limits_{0}^{\infty} \!\! u^{-\frac 12}
(p^2\!+\!u)^{\frac 12-i}(1\!+\!u)^{-\frac {1+\alpha}2}{\rm d}u, \quad 
(i=1,2).
\label{eq:aux-pi}
\end{equation}
These two integrals can be evaluated in terms of beta and 
hypergeometric 
functions (Gradshteyn \& Ryzhik 1980): 
\begin{equation} 
P_i(p,\alpha) = p^{2-2i}{\rm B}\left(
       {\textstyle \frac 12,\frac {2i\!-\!1\!+\!\alpha}2} \right)  
{}_2F_1\left( {\textstyle 
       \frac{1\!+\!\alpha}2,\frac 12;i\!+\!\frac 
{\alpha}2;1\!-\!p^2}\right).
\label{eq:explicit-aux-pi}
\end{equation}
They reduce to elliptic integrals when $\alpha=0, 1$.

A necessary condition for self-consistency is obtained from 
(\ref{eq:criterion-power-law}) and (\ref{eq:elliptic-discs-qs}) by
assuming $\lambda _{\rm min}=0$:
\begin{equation}
(1-p^2)\frac {P_2(p,\alpha)}{P_1(p,\alpha)}
< 1 + \frac {\alpha}2.
\label{eq:elliptic-discs-condition} 
\end{equation}  
The boundary curve between the allowed and forbidden zones of 
(\ref{eq:elliptic-discs-condition}) was computed numerically and is 
displayed in Figure~7 (solid line).  The admissible axial ratios of
the logarithmic discs have an infimum of $p=0.545$. This result agrees
with the findings of K93, whose numerical studies predicted a lower
bound of $p\approx0.8$ for self-consistent models. As in the case of
the power-law discs, the numerical solutions lie well inside the allowed
region of parameter space.

\begin{figure}
\centerline{\hbox{\epsfxsize=3.1in\epsfbox{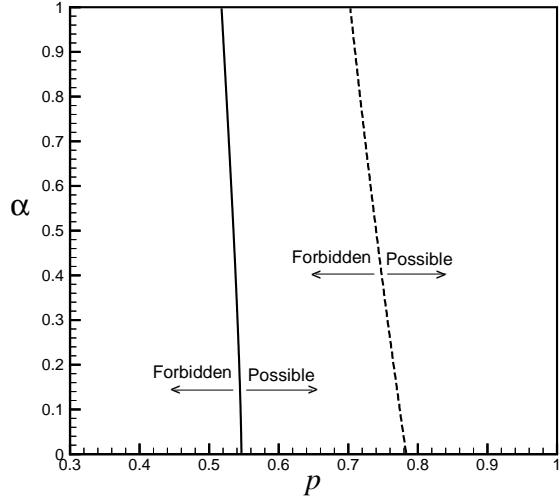}}}
\caption{The parameter space of elliptic discs and projected
power-law galaxies. The boundary curves between possible and
non-consistent models are shown by solid and dashed lines
for elliptic discs and projected power-law galaxies, respectively.
Non-consistent models are to the left of boundary curves.
The parameter $p$ is the axial ratio of isodensity contours of
elliptic discs. For the projected power-law galaxies, however,
the parameter $p$ is related to the axial ratio of the surface
density by eqs (B5), (B6) and (B11).}
\end{figure} 
 
\subsection{Projected power-law galaxies}
\label{sec:projected-discs}

Projection of the density of scale-free triaxial models with
potentials that are stratified on similar concentric ellipsoids
(triaxial versions of the power-law galaxies of Evans 1994) provides
elongated discs with surface-densities of the form
(\ref{eq:surf-density}) and associated potentials of the form
(\ref{eq:potential}). We have not found this potential-density pair in
the literature, and give the full derivation in Appendix B. The
function $S(\phi)$ is given in eq.~(\ref{eq:s-mu}), and the function
$P(\phi)$ is given in eqs (\ref{eq:ppd-potential-gen}) and
(\ref{eq:ppd-potential-spec}). We will refer to these discs as the
{\it projected power-law galaxies}.
 
On the major axis ($\phi=0$) of projected power-law galaxies, we
find
\begin{eqnarray}
Q_{\mu}  \!\!\!\! &=& \!\!\!\! \frac {3\alpha ^2 p^4+6-6\alpha + 
p^2(9\alpha\!-\!4\!-\!2\alpha ^2)}{p^2 (2\!+\!\alpha)(1\!+\!\alpha 
p^2)}, 
\nonumber \\ 
Q_V \!\!\!\! &=& 1+{1-p^2 \over p^2}{U_2(p,\alpha) \over
U_1(p,\alpha)}, 
\label{eq:ppd-curvatures}
\end{eqnarray}
where  
\begin{eqnarray}  
U_i(p,\alpha) \!\!\!\!\! &=& \!\!\!\!\! {\rm B}    
\left ({\textstyle \frac 12,\frac {2i\!+\!1\!+\!\alpha}2 }\right)
\times             \nonumber \\
\!\!\!\!\! &{}& \!\!\!\!\! \Big [  
p^2(1\!+\!\alpha) \, _2F_1 \left({\textstyle \frac 
{3\!+\!\alpha}2,\frac 12;  
i\!+\!1\!+\!\frac {\alpha}2;1\!-\!p^2 } \right) + \nonumber \\  
\!\!\!\!\! &\phantom{+}&  (2i\!-\!1) \, _2F_1 \left (  
{\textstyle \frac {1\!+\!\alpha}2,\frac 12;i\!+\!1\!+\!\frac {\alpha}2;  
1\!-\!p^2} \right ) \Big ],  
\label{eq:u-function-ppd}  
\end{eqnarray}  
with $i=1,2$. One can easily verify that $Q_{\mu}>0$ for all values of  
$\vert \alpha \vert \le 1$ and $0 \le p\le 1$. Therefore, the  
necessary condition for self-consistency becomes  
\begin{equation}  
{(1-p^2) \over p^2}{U_2(p,\alpha) \over U_1(p,\alpha)} <  
1+\frac {\alpha}2.   
\label{eq:ppd-curvature-condition}  
\end{equation}  
The boundary curve between the allowed and forbidden zones of  
inequality (\ref{eq:ppd-curvature-condition}) is displayed in  
Figure~7 (dashed line). In comparison with the power-law discs, 
the self-consistency of a larger fraction of the parameter 
space is ruled out for the projected power-law galaxies.   
  
\subsection{Sridhar--Touma discs}  
\label{sec:curvature-st-discs}  
  
The separable discs introduced by STa have  
\begin{equation}   
P(\phi) =\left [ (1+\sin \phi)^{1+\alpha}   
+(1-\sin \phi)^{1+\alpha} \right ], 
\label{eq:st97_1}   
\end{equation}   
where $0<\alpha <1$. The associated density function $S(\phi)$ is 
given in eq.~(4) of STa as a one-dimensional integral. From 
(\ref{eq:st97_1}) we obtain $Q_V=2+\alpha$ and $Q_{\mu}>0$. Upon 
substituting these results into (\ref{eq:criterion-power-law}) we 
arrive at the requirement 
\begin{equation}   
2+\alpha < 2+ \frac {\alpha}2,   
\label{eq:st97_2}   
\end{equation}   
which shows that self-consistent models are ruled out for all 
admissible values of $\alpha$. This is identical to the conclusion 
drawn by SZ98, based on another method. The non-self-consistency of 
STa models can also be interpreted by our arguments of 
\S\ref{sec:orbit-structure}. All STa models, which have the fixed 
axial ratio $p=0.5$ of equipotential curves (for all values of the 
cusp slope), are non-self-consistent because they host only a 
continuous family of centrophobic banana orbits.

\section{Scale-free axisymmetric galaxies} 
\label{sec:curvature-three} 
 
We now extend the curvature condition to scale-free axisymmetric 
models, in which the orbits can be described as two-dimensional motion 
in the meridional plane. 
 
\subsection{Potential-density pairs} 
 
Consider spherical polar coordinates $(r,\theta,\phi)$ where $r$ 
is the radial distance from the origin, $\theta$ is the colatitude and 
$\phi$ is the azimuthal angle. We denote the momenta conjugate to 
$(r,\theta,\phi)$ by $(p_r,p_{\theta},p_{\phi})$. In axisymmetric 
systems, the component $p_{\phi}\equiv L_z$ of the angular momentum 
vector parallel to the symmetry axis is a conserved quantity. 
Furthermore, the variable $\phi$ does not explicitly occur in the 
potential-density pairs, which take the following forms in scale-free 
models 
\begin{eqnarray} 
\label{eq:axisym-pot-dens} 
V \!\! &=& \!\! \left\{ \begin{array}{ll}  
         V_0 {\rm sgn}(\alpha)r^{\alpha}P(\theta), &\alpha\not=0, \\ 
         \null & \null \\ 
         V_0 [2\ln r +P(\theta)], &\alpha=0,  
         \end{array} \right. \nonumber \\ 
\rho \!\! &=& \!\! \rho_0 r^{\alpha -2}S(\theta),\qquad \alpha <2. 
\end{eqnarray} 

\subsection{Derivation of the curvature condition}
\label{sec:curvature-three-derivation}

The curvature condition (\ref{eq:criterion-one}) can be extended to
three-dimensional axisymmetric models if we define $\Gamma$ as
\begin{equation}
\Gamma =\frac {1}{\Vert {\boldmath L}_j \Vert \mu (r_j,\theta _j)},
\label{eq:axi-curvature}
\end{equation}
where
\begin{equation}
{\boldmath L}=\left ( p_{\theta}^2 +
        \frac {L_z^2}{\sin ^2 \theta} \right )^{1/2},
\label{eq:axi-angmom}
\end{equation}
is the modulus of the angular momentum and (cf.\ 
eq.~[\ref{eq:rj-gamma}])
\begin{equation}
\mu (r_j,\theta _j) =
r_j^{-\gamma}S(\theta _j), \qquad \gamma =1+\frac {\alpha}2.
\label{eq:mu-and-gamma}
\end{equation}
The subscript $j$ refers to the $j$th orbit family. The scaling
properties of planar discs given in (\ref{eq:scaling}) will be valid
for three-dimensional models if we replace $R$ with $r$. The
smoothness condition for $\Gamma$ becomes
\begin{equation}
\sum _{j=1}^{\infty}w(j) \left \langle \frac {{\rm d}^2 \Gamma}
{{\rm d}\theta _j^2} \right \rangle =0, \qquad\forall \theta _j
\in \left [0,\frac {\pi}{2}\right ],
\label{eq:axi-smoothness}
\end{equation}
where $\theta =\pi/2$ corresponds to the equatorial plane. This is
identical to equation (\ref{eq:second-derivative}), but now applies to
axisymmetric systems. As before, we drop the subscript $j$ and write
\begin{eqnarray}
\frac 1{\Gamma} \frac {{\rm d}^2\Gamma}{{\rm d}\theta ^2} \!\!\! &=& 
\!\!\!
-\frac 1{\boldmath L} \frac {{\rm d}^2 {\boldmath L}}{{\rm d}\theta ^2}
-\frac 1{\mu} \frac {{\rm d}^2 \mu}{{\rm d}\theta ^2} +
2 \frac {{\rm d}{\boldmath L}}{\boldmath L {\rm d}\theta}
\frac {{\rm d}\mu}{\mu {\rm d}\theta} \nonumber \\
\!\! &\phantom{=}& \!\! + 2 \left ( \frac 1{{\boldmath L}}
\frac {{\rm d}{\boldmath L}}{{\rm d}\theta} \right )^2+
2 \left ( \frac {1}{\mu} \frac {{\rm d}\mu}{{\rm d}\theta}
\right )^2.
\label{eq:axi-second-deriv}
\end{eqnarray}
We confine ourselves to the region near the equatorial plane. This
simplifies the expressions considerably. For the models having
reflection symmetry with respect to the equatorial plane, we have
$\partial \mu /\partial \theta$= $\partial V /\partial \theta$=0 at
$\theta =\pi /2$.

The first and second derivatives of ${\boldmath L}$ and $\mu$ with
respect to $\theta$ can be obtained using the equations of motion
\begin{equation}
\dot r=p_r, \qquad \ddot r=\dot p_r=-\frac {\partial V}{\partial r}
+\frac {p_{\theta}^2}{r^3}+\frac {L_z^2}{r^3 \sin ^2 \theta},
\label{eq:axi-motion-radial}
\end{equation}
\begin{equation}
\dot \theta =\frac {p_{\theta}}{r^2}, \qquad
\dot p_{\theta}=-\frac {\partial V}{\partial \theta}+
\frac {L_z^2\cos \theta}{r^2\sin ^3 \theta}.
\label{eq:axi-motion-angle}
\end{equation}
At $\theta =\pi /2$, eq.~(\ref{eq:axi-motion-angle}) implies $\dot
p_{\theta}={\rm d}p_\theta/{\rm d}\theta \approx 0$, which leads to
$r\ddot \theta +2 \dot r \dot \theta \approx 0$ near the equatorial
plane. Furthermore, from (\ref{eq:axi-angmom}) we find that ${\rm
d}{\boldmath L}/{\rm d}\theta$ vanishes at $\theta=\pi/2$.  By using
(\ref{eq:axi-motion-angle}) we find
\begin{equation}
\left [-\frac {1}{{\boldmath L}}\frac {{\rm d}^2{\boldmath L}}
{{\rm d}\theta ^2}\right ]_{\theta =\frac {\pi}2} = \left [ \frac {1}
{1+\frac {L_z^2}{p_{\theta}^2}} \left ( \frac 1{r^2\dot \theta ^2}
\frac {\partial ^2V}{\partial \theta ^2} \right ) \right ]_
{\theta = \frac {\pi}2}.
\label{eq:axi-second-deriv-l}
\end{equation}
The derivatives of $\mu$ with respect to $\theta$ can be expressed in
terms of ${\rm d}r/{\rm d}\theta$ and ${\rm d}^2r/{\rm d}\theta ^2$,
which we evaluate by means of the equations of motion and the
condition $r\ddot \theta +2 \dot r \dot \theta \approx 0$ which is
valid near $\theta=\pi/2$.

Combining all terms in expression (\ref{eq:axi-second-deriv}), and
carrying out some algebraic manipulations, then leads to
\begin{equation}
\left [ \frac {{\rm d}^2\Gamma}{{\rm d}\theta ^2}
\right ]_{\theta =\frac {\pi}2} = \left \{ \frac {\Gamma}{K_{\theta}}
\left [ \kappa Q_V - (\kappa +\gamma -\gamma \lambda) \right ]
\right \}_{\theta =\frac {\pi}2},
\label{eq:axi-condition-one}
\end{equation}
where
\begin{equation}
\lambda = Q_{\mu} K_{\theta} +(1+\gamma)K_r +K_{\phi},
\label{eq:axi-condition-two}
\end{equation}
with
\begin{equation}
K_r     = \frac {\dot r ^2}{r\frac {\partial V}{\partial r}},
\quad K_{\theta}=\frac {(r\dot \theta)^2}{r\frac {\partial V} 
{\partial r}}, \quad K_{\phi}=\frac {1}{r\frac {\partial V} 
{\partial r}}\frac {L_z^2}{r^2}, 
\label{eq:axi-condition-three} 
\end{equation} 
and 
\begin{equation} 
\kappa = \left ( 1 +\frac {K_{\phi}}{K_{\theta}} \right )^{-1}, 
\label{eq:axi-condition-four} 
\end{equation} 
and the quantities $Q_V$ and $Q_{\mu}$ are defined by 
\begin{equation} 
Q_V=1+\frac {\frac {\partial ^2V}{\partial \theta ^2} } 
{r \frac {\partial V}{\partial r}}|_{\theta =\frac {\pi}2}, 
\qquad 
Q_\mu=1+\frac {\frac {\partial ^2\mu}{\partial \theta ^2} } 
{r \frac {\partial \mu}{\partial r}}|_{\theta =\frac {\pi}2}. 
\label{eq:axi-condition-five} 
\end{equation} 
These expressions resemble those for the elongated disc case, but 
contain the additional parameter $\kappa$.  
 
A necessary condition for self-consistency follows from 
(\ref{eq:axi-condition-one}) as 
\begin{equation} 
\kappa +\gamma -\gamma \lambda _{{\rm max}} < 
\kappa Q_V < \kappa +\gamma -\gamma \lambda _{{\rm min}}. 
\label{eq:axi-condition-six} 
\end{equation} 
The greatest upper bound of (\ref{eq:axi-condition-six}) corresponds 
to $\lambda _{\rm min} =0$. Therefore, a non-trivial condition for 
self-consistency is\looseness=-2 
\begin{equation} 
Q_V<1+\frac {\gamma}{\kappa}. 
\label{eq:axi-condition-seven} 
\end{equation} 

\section{Applications} 
\label{sec:app-oblate} 
 
We now apply the curvature criterion to three families of scale-free 
axisymmetric galaxy models. 
 
\subsection{The power-law galaxies} 
\label{sec:curvature-three-power-law} 
The power-law galaxies of Evans (1993, 1994) have 
\begin{equation} 
S(\theta) \!=\! \left\{ \begin{array}{ll}  
            \!\!\! \alpha \! \left[ m_1^2(\theta) \right 
]^{\frac{\alpha -4}{2}}  
            \!\! \left[ 2q^2 \!-\! m_1^2(\theta) \!+\! \alpha 
m_2^2(\theta)   
            \right],\!\!\! &\alpha\not=0, \\ 
            \null & \null \\ 
            \!\!\! 2 \! \left [ m_1^2(\theta) \right ]^{-2} 
            \left [ 2q^2-m_1^2(\theta) \right ],\!\!\!  &\alpha=0,  
            \end{array} \right. 
\end{equation} 
and  
\begin{equation} 
P(\theta) \!=\! \left\{ \begin{array}{ll}  
             \! \left [m_1^2(\theta) \right ]^{\alpha /2}, 
&\alpha\not=0, \\ 
             \null & \null \\ 
             \! \ln m_1^2(\theta),  &\alpha=0,  
	     \end{array} \right. 
\label{eq:power-law-pot-den} 
\end{equation}  
where $V_0=4\pi G \rho_0$ and we have defined  
$m_k^2(\theta)=q^{2k}\sin^2\theta+\cos^2\theta$.  
The density function $\rho$ is positive for $q^2>(1\!-\!\alpha)/2$.   
The parameter $\beta$ introduced in Evans (1994) is equivalent to 
$-\alpha$. 
 
From (\ref{eq:power-law-pot-den}) we find $Q_V=1/q^2$. Therefore, the 
condition (\ref{eq:axi-condition-seven}) reduces to 
\begin{equation} 
q^2>  \frac{2\kappa}{2\kappa+2+\alpha}. 
\label{eq:axi-power-law-condition} 
\end{equation} 
This shows that the self-consistency of the power-law models is not 
ruled out near the equatorial plane. 
 
The above inequality has an interesting interpretation: since the 
quantity $\kappa$ is a dynamical property of orbit families, as it 
involves not only the potential but also $L_z$, we can investigate 
whether an orbit family is useful for the self-consistency of a given 
model. For example, $\kappa=1$ corresponds to orbits confined 
completely to a meridional plane ($L_z=0$), and therefore, we can 
rule out self-consistent models with 
\begin{equation} 
q < \frac 1{\sqrt {2+\frac {\alpha}2} }, 
\label{eq:axi-condition-alt} 
\end{equation} 
using only the meridional orbits. This is the same result obtained for 
the power-law discs in \S\ref{sec:curvature-condition-power-law}.  
As $\kappa$ tends to zero, the amplitude of vertical motions decreases  
and more flattened models are supported. In the limit, the models with  
$q=0$ become self-consistent for $\kappa =0$. Thus, equatorial orbits  
are essential building blocks of highly flattened models in general  
and limiting circular discs in particular. 
 
\subsection{Scale-free spheroids} 
\label{sec:curvature-three-spheroids} 
 
Scale-free galaxy models with spheroidal density isocontours have 
widely been used in galactic dynamics (Qian et al.\ 1995, hereafter 
Q95). In this case  
\begin{equation} 
S(\theta) = \left [ m_1^2(\theta) \right ] ^{\frac{\alpha}{2} -1},  
\end{equation} 
and 
\begin{equation} 
P(\theta)\!=\!\!\left\{ \begin{array}{ll}  
               \!\!\! \int\limits _{0}^{\infty}  
{\displaystyle \frac{(q^2\sin^2\theta +  
               \cos^2 \theta +u)^{\frac {\alpha}2} {\rm d}u} 
               {(1\!+\!u)^{1+\frac{\alpha}2}(q^2\!+\!u)^ 
{\frac {1+\alpha}2}} }, &\!\!\alpha\not\!=\!0, \nonumber \\ 
                 \null & \null \\ 
                \!\!\! {\pi G\rho_0 \over q V_0 }\int\limits 
_{0}^{\infty}   
{ \displaystyle      
{  \ln (q^2\sin^2\theta +\cos^2\theta +u) {\rm d}u \over  
    (1+u)(q^2\!+\!u)^{\frac 12} } },\!\! 
   &\alpha\!=\!0,  
		 \end{array} \right.  
\label{eq75} 
\end{equation} 
where $V_0=2\pi G \rho_0 q^{\alpha -1}/\alpha$ for $\alpha \not =0$ and  
$V_0=2\pi G \rho_0 R_F(1,1,q^2)/q$ for $\alpha =0$. $R_F$ is the  
Carlson elliptic integral of the first kind. 
 
Oblate models have $0<q<1$ and prolate models have $q>1$. We restrict  
ourselves to the range $-1 < \alpha < 1$. 
 
The curvature parameter $Q_V$ (in the equatorial plane $\theta 
=\pi/2$) is 
\begin{equation} 
Q_V = 1+(1-q^2)\frac {V_2(q,\alpha)}{V_1(q,\alpha)}, 
\label{eq76} 
\end{equation} 
where 
\begin{equation} 
V_i(q,\alpha) \!=\! q^{2\!-\!2i}{\rm B}{\textstyle \left (1,i\!+\!\frac 
                       {\alpha -1}2 \right)}  
{}_2F_1 {\textstyle \left(1\!+\! 
         \frac{\alpha}2,1;i\!+\!\frac{1\!+\!\alpha}2; 1\!-\!q^2 
\right)}, 
\label{eq77.1} 
\end{equation} 
for oblate models and 
\begin{equation} 
V_i(q,\alpha)\!= \!q^{2\!-\!2i}{\rm B} {\textstyle \left (1,i\!+\! 
             \frac {\alpha\! -\!1}2 \right) }  
{}_2F_1{\textstyle \left(\frac {2i\!-\!1}2,1;i\!+\!\frac 
{1\!+\!\alpha}2; 
1\!-\!\frac 1{q^2} \right)}, 
\label{eq77.2} 
\end{equation} 
for prolate ones. By inserting expression (\ref{eq76}) in the 
condition (\ref{eq:axi-condition-seven}), and noting that $\gamma 
=1+\frac {\alpha}2$, we obtain the necessary condition for 
self-consistency as 
\begin{equation} 
\frac {(1-q^2)V_2(q,\alpha)}{V_1(q,\alpha)}<\frac 1{\kappa} 
\left ( 1+\frac {\alpha}2 \right ). 
\label{eq78} 
\end{equation} 
Numerical computations show that the left-hand side of (\ref{eq78}) is 
positive for oblate models and negative for prolate ones. So, the 
condition (\ref{eq78}) is fulfilled by the proper choices of 
$0<\kappa<1$, and we cannot rule out any models.

\subsection{Oblate Sridhar--Touma models} 
\label{sec:curvature-three-st} 
 
The oblate mass models described by Sridhar \& Touma (1997b, hereafter 
STb) are the extensions of the flat STa models to three-dimensional 
space. These axisymmetric, cuspy models are integrable and admit an 
exact third integral of motion. We have shown elsewhere (Jalali \& de 
Zeeuw 2001) that exact two- and three-integral distribution functions 
can be constructed for these models. It is useful to determine what 
the curvature condition gives. 
 
The STb models have 
\begin{equation} 
S(\theta) = (\alpha\! -\!\cos \theta) 
         (1\!+\!\cos \theta)^{\alpha} 
       \!+\! (\alpha\! +\!\cos \theta)( 
       1\!-\!\cos \theta)^{\alpha},  
\end{equation} 
and 
\begin{equation} 
P(\theta) = (1\!+\!\cos \theta)^{1+\alpha} \!+\! 
      (1\!-\!\cos \theta)^{1+\alpha},  
\label{eq70} 
\end{equation} 
where $1<\alpha <2$. It follows that $Q_V$ is given by 
$Q_V=1+\alpha$, which upon substitution in 
(\ref{eq:axi-condition-seven}) yields the condition 
\begin{equation} 
\alpha < \frac 1{\kappa}\left ( 1+\frac {\alpha}2 \right ). 
\label{eq72} 
\end{equation} 
This is satisfied for $\alpha <2$ and all possible orbits 
corresponding to $0<\kappa <1$, as expected based on the results of 
Jalali \& de Zeeuw (2001). The phase space of STb models consists of a 
continuous family of reflected banana orbits all of which are useful 
for the construction of self-consistent equilibria.\looseness=-2

\section{Discussion}
\label{sec:discussion}

Lynden-Bell (1962) showed that the distribution function of a stellar
system depends on the phase-space coordinates through the isolating
integrals of motion. However, such integrals of motion are not
available for most physical models.  In fact, there are few
mathematical models of galaxies that have separable potentials.
Finding the integrals of motion is a formidable task for
non-integrable systems. The existence of $f(E, L_z)$ distribution
functions for axisymmetric models (with $E$ the orbital energy and
$L_z$ the angular momentum parallel to the symmetry axis) has helped
galactic dynamicists to develop more realistic galaxy models and
extract their observable properties (e.g., Q95). But, in
non-axisymmetric systems, only $E$ is conserved, and we have to adopt
other construction methods. Schwarzschild's (1979) paper was a major
step in this way. His elegant numerical method allowed astronomers to
attack the self-consistency problem of galaxy models from a different
point of view, by constructing the coarse-grained distribution
functions as weighted sums of the densities of individual stellar
orbits without explicit knowledge of non-classical integrals of
motion.

Schwarzschild's method does not require restrictions on the shape of
the model or the functional form of the potential, and can be applied
to non-integrable models as well as integrable ones. The only drawback
is the cost of the numerical computations, especially when the number
of cells in the configuration space is large and the orbit library is
rich. Although supercomputers can handle massive calculations, a quick
mathematical tool has always been attractive. ZCZ developed one such
tool, which is a simple mathematical test for self-consistency, and
applied it to scale-free elongated discs. The idea derives from the
fact that by increasing the number of orbits and cells in
Schwarzschild's method, one can obtain a limiting analytical
expression for the conditions of dynamical equilibrium.  The
smoothness condition for the quantity $\Gamma$ is such an
interpretation of Schwarzschild's method.  The curvature condition
says that if the discrete numerical equations (subject to the 
positivity constraints on the weight functions) is solvable, a 
relation between the curvatures of the potential and density functions 
must be satisfied. This results in a necessary condition for 
equilibrium.  It takes a particularly simple form near the major axis 
of the model where the surface density isocontours are strongly 
curved, and the reconstruction of the density profile is most 
difficult, as it requires boxlets and chaotic orbits. We have 
corrected an error in the ZCZ derivation of the curvature condition, 
and also extended it to motion in axisymmetric 
potentials.\looseness=-2 
   
The curvature condition is necessary but not sufficient, so we can use
it to detect non-consistency, but cannot prove self-consistency. We
can only state that in the allowed region of the parameter space,
explored by the curvature condition, scale-free self-consistent models 
may be found. Our experiments on the power-law discs show that the 
results of the curvature condition (applied to the major axis of 
elongated discs) do indeed provide a boundary inside which the numerical 
solutions lie, and they furthermore have highlighted the need to be 
careful with the choice of the catalog of zero-angular momentum orbits 
when applying Schwarzschild's method.
  
Our results reveal a correlation between the non-self-consistency 
index $Y$ and the relative population of banana orbits $M_{B/2}$ (as 
suspected in many of the earlier studies quoted). Maximally 
non-self-consistent power-law discs (for $p\approx 0.5$) have 
banana-rich orbital structures independent of the value of 
$\alpha$. We speculate that this property is shared by all razor-thin 
discs whose potential functions allow for the existence of 1:2 
resonant orbits, and conclude that it is of interest to search for 
more elongated discs without bananas (Jalali 1999; Sridhar \& Touma 
1999). 
  
In the case of three-dimensional axisymmetric models, we have employed 
our test near the equatorial plane where the circular orbits and 
equatorial rosettes overwhelm the effect of other orbit families. The 
existence of the $L_z$ integral provides an extra quantity $\kappa$ 
that can be controlled to satisfy the necessary condition of 
self-consistency.  This dynamical property of orbit families does not 
occur in the study of planar non-axisymmetric systems and is not 
expected to appear in triaxial systems. It is known that $f(E, L_z)$ 
distribution functions cannot be constructed for most prolate models 
(an upper bound exists for the allowed axial ratios of prolate models 
versus the cusp slope, e.g., Evans 1994, Q95), but the curvature 
condition does not reject the self-consistency of prolate galaxies. 
This is not a shortcoming, for the curvature condition is necessary 
but not sufficient.

Our next goal is to extend the curvature condition to scale-free
triaxial galaxies. The step to non-scale-free models remains as
another challenging problem.

\section{acknowledgments}
The authors thank HongSheng Zhao for useful discussions and
Nicolas Cretton for providing the RK78 routine. Comments by the
referee C. Siopis helped us to improve the presentation in \S3.
MAJ thanks the hospitality of Sterrewacht Leiden where this 
work was begun. NOVA, the Netherlands Research School for Astronomy, 
provided partial funding for the visit.

\appendix 

\section{The surface density of the power-law discs}
\label{sec:surf-den-power}

We derive the surface density associated with the disc potential
\begin{equation}
V(R,\phi) = V_0 R^\alpha (p^2\cos ^2 \phi+\sin ^2 \phi)^
{\alpha /2}, \qquad 0<\alpha <1,
\end{equation}
with $0<p<1$. We introduce $\epsilon=1-p$ and rewrite the disc
potential in the form
\begin{equation}
V=V_0 R^\alpha (1-2 t\cos \phi +t^2)^{\alpha / 2},
\end{equation}
where $t=\epsilon \cos \phi$ with $\vert t \vert <1$ (we exclude
the case $\epsilon =1$ from our calculations). The $\phi$-dependence
of $V$ can now be expanded in terms of Gegenbauer polynomials
as (see eq.~[8.930] in Gradshteyn \& Ryzhik 1980)
\begin{equation}
V=V_0 R^\alpha \sum_{n=0}^{\infty}
C^{\lambda}_n(\cos \phi)t^n,
\end{equation}
where $\lambda =-\alpha/2$ and
\begin{equation}
\label{eq:gegenbauer-cosine-series}
C^{\lambda}_n(\cos \phi) \!= \! \sum_{k=0}^{n}
{\Gamma (\lambda+k)\Gamma (\lambda +n-k) \over k! (n-k)!
[\Gamma (\lambda)]^2 } \cos (2k-n)\phi.
\end{equation}
We now expand the term $t^n$ in a Fourier series, and write
\begin{equation}
\label{eq::powers-of-cosine}
\cos ^n \phi=\sum_{m=0}^{n} \beta^{(n)}_m \cos m\phi,
\end{equation}
where $\beta^{(n)}_m=0$ when $n-m$ is odd and
\begin{eqnarray}
\beta^{(n)}_m \!\! &=& \!\! \left\{ 
\begin{array}{ll} \frac 1{2^{n-1}} \! \left ( \!\!\!\!
                          \begin{array}{c}  n \\
                                        {\displaystyle \frac{n-m}2}
                          \end{array} \!\!\!\! \right ),  &m \not=0, \\
                  \null & \null \\
                  \frac 1{2^n} \! \left ( \!\!\!\!
                          \begin{array}{c} n \\
                                       {\displaystyle     n/2}
                          \end{array} \!\!\!\! \right ), & m=0,  \\
\end{array} \right.
\label{eq:beta-potential}
\end{eqnarray}
when $n-m$ is even. Using expressions
(\ref{eq:gegenbauer-cosine-series}) and (\ref{eq::powers-of-cosine}),
the potential function becomes
\begin{eqnarray}
V \!\! &=& \!\! {V_0 R^\alpha\over 2} \sum_{n=0}^{\infty}\sum_{m=0}^n
\sum_{k=0}^n \epsilon ^n \beta^{(n)} _m
{\Gamma (\lambda+k)\Gamma (\lambda +n-k) \over k! (n-k)!
[\Gamma (\lambda)]^2 } \nonumber \\
&{}& \times [\cos (2k-n+m)\phi+\cos (2k-n-m)\phi].
\end{eqnarray}
Now, we use eqs (A1) and (A5) of Syer \& Tremaine (1996, hereafter 
ST96) to find the surface densities corresponding to single Fourier modes
and superpose the results. Interestingly, the surface density becomes
\begin{eqnarray}
\label{eq::power-law-density}
\Sigma \!\!\!\!\! &=& \!\!\!\!\! \frac {-V_0 R^{\alpha -1}}{2\pi G}
\sum_{n=0}^{\infty}
\sum_{m=0}^{n}\sum_{k=0}^{n} \epsilon ^n \beta^{(n)} _m
{\Gamma (\lambda+k)\Gamma (\lambda +n-k) \over k! (n-k)!
[\Gamma (\lambda)]^2 } \nonumber \\
&{}& \times [F(s_1)\cos s_1\phi+F(s_2)\cos s_2\phi],
\end{eqnarray}
where
\begin{eqnarray}
F(s_i) \!\!\! &=& \!\!\! 
{\Gamma(s_i/2+\alpha/2+1)\Gamma(s_i/2-\alpha/2+1/2)
\over \Gamma(s_i/2+\alpha/2+1/2) \Gamma(s_i/2-\alpha/2)}, \nonumber \\
s_1 \!\!\! &=& \!\!\! 2k-n+m, \nonumber \\
s_2 \!\!\! &=& \!\!\! 2k-n-m.
\end{eqnarray}
The minus sign in front of (\ref{eq::power-law-density}) is
cancelled once the Gamma functions are evaluated. 

The case $\alpha =0$, which corresponds to logarithmic potentials, is
special and needs a different treatment. In this case we have
\begin{equation}
\label{eq:pot-log}
V(R,\phi)=V_0 \left [ 2\ln R + \ln (1 - 2t \cos \phi +t^2) \right ].
\end{equation}
The second term in (\ref{eq:pot-log}) can be expanded as
the series (Gradshteyn \& Ryzhik 1980, eq. [1.514])
\begin{equation}
\ln (1 - 2t \cos \phi +t^2) =
-2\sum_{n=1}^{\infty} {\cos n \phi \over n}t^n,
\end{equation}
which after the substitution $t=\epsilon \cos \phi$ and
using $\ln (1-z)=-\sum_{n=1}^{\infty}z^n/n$, gives the
potential as
\begin{eqnarray}
V \!\!\!\! &=& \!\!\!\! V_0 \Big \{2\ln \Big (1-{\epsilon \over 2} 
\Big ) + 2\ln R -\sum_{n=1}^{\infty}{\epsilon ^n
\over 2^{n-1} n} \cos 2n\phi \nonumber \\
\!\!\!\! &-& \!\!\!\! \sum_{n=1}^{\infty} \sum_{m=0}^{n-1}
{\epsilon ^n \beta ^{(n)}_m \over n} \left [\cos (n+m)\phi
+ \cos (n-m)\phi \right ] \Big \}.
\end{eqnarray}
By dropping the constant term in $V$ and using eq.\ (A5) of ST96, we
obtain the surface density as
\begin{eqnarray}
\Sigma(R,\phi) \!\!\!\! &=& \!\!\!\! \frac {V_0}{\pi G R}
\Big [ 1 + \sum_{n=1}^{\infty} {\epsilon ^n \over 2^{n-1}}
\cos k\phi \nonumber \\
\!\!\!\! &+& \!\!\!\! \sum_{n=1}^{\infty}
\sum_{m=0}^{n-1} {\epsilon ^n \beta ^{(n)}_m
\over 2n} ( i \cos i\phi + j \cos j\phi ) \Big ],
\label{eq:explicit-surf-den}
\end{eqnarray}
with $k=2n$, $i=n+m$ and $j=n-m$. 

A simple representation for $\Sigma$ is
\begin{equation}
\label{eq:power-law-density-fourier}
\Sigma(R, \phi) = {1\over R} \Big ( a_0+\sum_{\ell=1}^{\infty}
a_{\ell}\cos \ell \phi \Big ),~~a_0={V_0\over \pi G},
\end{equation}
where the Fourier coefficients $a_{\ell}$ ($\ell \ge 1$)
are given by
\begin{equation}
a_{\ell} = {V_0\over \pi G}\sum_{n=1}^{\infty} \left [
{\epsilon ^n \delta _{k,\ell} \over 2^{n-1}}+\sum_{m=0}^{n-1}
{\epsilon ^n \beta ^{(n)}_m \over 2n}
(i\delta _{i,\ell}+j\delta _{j,\ell}) \right ],
\end{equation}
with
\begin{eqnarray}
\delta _{r,s} \!\! &=& \!\!
\left\{ 
\begin{array}{ll} 1,  &r=s, \\
                  0,  &r\not =s.  \\
\end{array} \right.
\end{eqnarray}
From the definition of $\beta ^{(n)}_m$, we conclude that $a_{\ell}=0$
when $\ell$ is odd. Using the rule $\Gamma (z+1)=z\Gamma (z)$ and eq.\
(A5) of ST96, the potential function can be regenerated from
(\ref{eq:power-law-density-fourier}) as
\begin{equation}
V=2\pi G a_0 \ln R -2\pi G \sum_{{\rm even}~\ell > 0}^{\infty}
{a_{\ell} \over \ell} \cos \ell \phi,
\end{equation}
which is identical to the result obtained by K93 for aligned discs
(the coefficients $a_{\ell}$ of K93 are computed for an elliptic
disc with flat rotation curve).

\section{The projected power-law galaxies} 
\label{sec:projected-power} 

We compute the projected surface density of the scale-free triaxial 
power-law models, for arbitrary direction of viewing, and show that 
the resulting surface density has a simple form. We consider it as the 
surface density of an elongated scale-free disc, and show that the 
gravitational potential in this disc can be evaluated explicitly. This 
then defines a new family of elongated discs which we call the 
projected power-law galaxies.

\subsection{Scale-free triaxial power-law models} 
  
Consider scale-free triaxial power-law models with gravitational 
potential 
\begin{equation}  
\label{eq:triaxial-potential}  
V(x,y,z)=\left\{ \begin{array}{ll}  
                 {\scriptsize \frac 12} v_c^2  
                 {\displaystyle  
                 \ln \Big (x^2+{y^2\over p^2}+ 
                 {z^2\over q^2} \Big ) }, &\alpha=0, \\  
                 \null & \null \\ 
		 \frac 1{\alpha}v_c^2  
                 {\displaystyle \Big (x^2+{y^2\over p^2}+  
		 {z^2\over q^2} \Big )^{\alpha/2}}, &0<\alpha<1.  
		 \end{array} \right. 
\end{equation}  
The associated density distribution is (Binney 1981; de Zeeuw \&  
Pfenniger 1988)  
\begin{equation}  
\label{eq:triaxial-density}  
\rho(x,y,z)={v_c^2 \over 4\pi G}{Ax^2+By^2+Cz^2 \over  
(x^2+p^{-2}y^2+q^{-2}z^2)^{2-\frac{\alpha}{2}}},  
\end{equation}  
with  
\begin{eqnarray}  
A \!\!\! &=& \!\!\! {1 \over p^2}+{1 \over q^2}-1+\alpha, \nonumber \\  
B \!\!\! &=& \!\!\! {1 \over p^2} \left( 1-{1-\alpha \over p^2}+  
                     {1 \over q^2} \right), \\  
C \!\!\! &=& \!\!\! {1 \over q^2} \left( 1+{1 \over p^2}-{1-\alpha 
\over q^2}  
                     \right). \nonumber  
\end{eqnarray}  
This is the triaxial generalization of the axisymmetric power-law
galaxies of Evans (1993, 1994). The surfaces of constant density are
not ellipsoidal and become dimpled near the short axis when $q$ is
small (e.g., Evans, Carollo \& de Zeeuw 2000). The density
distribution (\ref{eq:triaxial-density}) is non-negative as long as
$q^2+p^2q^2-(1-\alpha)p^2 \ge 0$.  This is a non-trivial constraint
except in the limit $\alpha \rightarrow 1$.

\subsection{Projected surface density} 

Now observe the density (\ref{eq:triaxial-density}) from an arbitrary
direction defined by the standard spherical polar coordinates
$(\theta, \phi)$. Define a coordinate system $(x'',y'',z'')$ such that
the $z''$ axis lies along the line-of-sight and the $x''$ axis lies in
the $(x, y)$-plane. The relations between $(x, y, z)$ and 
$(x'',y'',z'')$ 
are given in, e.g., de Zeeuw \& Franx (1989). Direct integration
shows that $\Sigma (x'',y'')$ is an elementary function, given by
\begin{eqnarray} 
\label{eq:projected-surface-density}  
\Sigma (x'',y'') \!\!\!\!\! &=& \!\!\!\!\! {v_c^2 \over 4\pi G}  
{ pq {\rm B} \left ( \frac 12,\frac 12\! -\!\frac \alpha 2 \right ) 
\over  
   \left ( c_1c_3-\frac 14 c_2^2 \right )^{1+\frac{\alpha}{2} }}  
                                                   \times \nonumber \\  
\!\!\!\!\! &\phantom{=}& \!\!\!\!\! { a_1 {x''}^2 - a_2 {x''}{y''}+  
a_3 {y''}^2 \over (c_1 {x''}^2-c_2 {x''}{y''}  
             +c_3{y''}^2 )^{\frac{3}{2}-\frac{\alpha}{2}}},  
\end{eqnarray}  
where  
\begin{eqnarray} 
c_1 \!\! &=& \!\! \sin ^2 \phi + p^2 \cos ^2 \phi, \nonumber \\
c_2 \!\! &=& \!\! 2(1-p^2)\sin \phi \cos \phi \cos \theta, \\ 
c_3 \!\! &=& \!\! \cos ^2 \phi \cos ^2 \theta 
       +p^2 \sin ^2 \phi \cos ^2 \theta +q^2 \sin ^2 \theta, \nonumber
\end{eqnarray}
and 
\begin{eqnarray} 
a_1 \!\! &=& \!\! c_1c_3 -{\textstyle\frac 14} c_2^2 
+\alpha \left ( c_1^2 + {\textstyle\frac 14} c_2^2 \right ), \nonumber 
\\ 
a_2 \!\! &=& \!\! \alpha c_2(c_1+c_3), \\ 
a_3 \!\! &=& \!\! c_1c_3 - {\textstyle\frac 14} c_2^2 
+\alpha \left ( c_3^2 +{\textstyle\frac 14} c_2^2 \right ). \nonumber 
\end{eqnarray} 
The surface density is scale-free, and has biaxial symmetry.

Now transform to polar coordinates $(R, \Phi'')$ defined by $(x'',y'')
=(R\cos\Phi'', R\sin\Phi'')$. Then the major axis of $\Sigma$ lies at
position angle $\Phi''=\Phi^*$, with 
\begin{equation}  
\tan 2\Phi^*={c_2 \over c_3-c_1} = {a_2 \over a_3-a_1}.   
\end{equation}  
This is the expected result: as the potential is stratified on 
ellipsoids, its projection is stratified on similar concentric 
ellipses at the fixed position angle $\Phi^*$ (Stark 1977), and this 
is then also the position angle of the projected surface density.  
 
When we rotate the coordinate system to $(R, \Phi)$ where 
$\Phi=\Phi''-\Phi^*$, we obtain 
\begin{equation}  
\Sigma (R,\Phi)=\Sigma _0 R^{\alpha-1} S(\bar \mu),  
\end{equation}  
where  
\begin{equation}  
\label{eq:sigma-0}  
\Sigma _0={v_c^2 \over 4 \pi G}  
\frac { 2^{\frac{3}{2}+\frac{\alpha}{2}}  
       {\rm B} \left(\frac 12,\frac{1-\alpha}2 \right)  
         pqP} {(c_1+c_3-\Delta)^{\frac{3}{2}+\frac{\alpha}{2}}},  
\end{equation}  
and  
\begin{equation}  
\label{eq:s-mu}  
S(\bar \mu)={(1-\alpha)P^2+\alpha(1+P^2) 
\bar \mu \over \bar \mu^{\frac{3}{2}-\frac{\alpha}{2}}}.  
\end{equation}  
with  
\begin{equation}  
P^2={c_1+c_3 -\Delta \over c_1+c_3 +\Delta}, \qquad  
  \Delta=\sqrt{(c_1-c_3)^2+c_2^2}  
\end{equation}  
and we have defined the abbreviation $\bar \mu=P^2\cos^2\Phi +\sin 
^2\Phi$ 
which will be useful below. 
 
Unlike $\rho$, the projected surface density $\Sigma$ is everywhere 
positive. The angular dependence of the projected surface density has 
the same form for {\it all} directions of observation, but the value of 
$P$ is a function of the intrinsic axis ratios $p$ and $q$ and the 
viewing angles $\theta$ and $\phi$. The isophotes are slightly 
oval. An example is shown in Figure 4 of Evans, Carollo \& de Zeeuw 
(2000).  The axis ratio $P'$ of the surface density, defined as 
$S(x',0)=S(0,P'x')$ is given by 
\begin{equation}   
P' ={P \over [(1+\alpha)(1+\alpha p^2)]^{1/(1-\alpha)}}  
\end{equation}  
which reduces to $P$ when $\alpha=0$. When the model is observed down 
the $z$-axis, we have $\phi$ is arbitrary, $\theta=0$, and $P=p$. 
 
\subsection{Potential in the disc}

Now consider surface densities of the form
(\ref{eq:projected-surface-density}) with $S(\bar \mu)$ given by
(\ref{eq:s-mu}) where without loss of generality we take $P=p$ and
$\Phi=\phi$ (see above).  We take $\Sigma_0$ given, but note that
(\ref{eq:sigma-0}) could be used to relate it to a three-dimensional
model.
 
The potential $V(x,y)$ in the plane of the disc can be evaluated with
the formalism developed by Evans \& de Zeeuw (1992, \S5). They
decompose a given biaxial scale-free disc into a weighted integral of
constituent discs with elliptic surface densities of different axis
ratios.  As the potential of the constituent discs is known from
classical theory, their weighted integral provides the potential of
our given disc.  Specifically, if we can find the function $w(\tau)$
such that
\begin{equation} 
S(\bar \mu)=\int\limits_{0}^{\infty} {w(\tau) {\rm d}\tau
                  \over (\tau +\bar 
\mu)^{\frac{1}{2}-\frac{\alpha}{2}}}, 
\label{eq:integral-equation}
\end{equation} 
then the potential in the disc is 
\begin{eqnarray} 
V(R,\bar \mu)\!\!\!\!\! &=& \!\!\!\!\! 
             {2\pi G \Sigma _0 R^{\alpha} \over 
             \alpha {\rm B}\left(\frac12, 
\frac12\!-\!\frac{\alpha}2\right)} 
      \!\! \int\limits_{0}^{\infty} \! w(\tau)[(1\!+\!\tau) 
                  (p^2\!+\!\tau)]^\frac{\alpha}{2} {\rm d}\tau 
\nonumber \\ 
\!\!\!\!\! &\times & \!\!\!\!\!\! \int\limits_{0}^{\infty} \!\!\!\! 
{(u\!+\!\bar \mu\!+\!\tau)^{\frac{\alpha}{2}} {\rm d}u \over 
  u^{1/2} [(u\!+\!p^2\!+\!\tau) 
         (u\!+\!1\!+\!\tau)]^{\frac{1}{2}+\frac{\alpha}{2}}},
\label{eq:disc-potential-alpha-not-0} 
\end{eqnarray} 
when $\alpha\not=0$, and 
\begin{equation} 
\label{eq:disc-potential-alpha-0} 
V(R,\bar \mu)=A \ln R +B(\bar \mu), 
\end{equation} 
with 
\begin{eqnarray} 
A \!\!\!\!\! &=& \!\!\!\!\! 4G\Sigma _0 \!\int\limits_{0}^{\infty}\! 
         w(\tau) R_F(\tau +1,\tau +p^2,0) {\rm d}\tau, \nonumber \\ 
B(\bar \mu) \!\!\!\!\! &=& \!\!\!\!\! G \Sigma_0 
\int\limits_{0}^{\infty} \!\! 
\int\limits_{0}^{\infty} \!\! {\ln (u\!+\!\bar \mu\!+\!\tau) w(\tau) 
{\rm d}\tau 
{\rm d}u \over u^{1/2} (u\!+\!p^2\!+\!\tau)^{1/2} 
         (u\!+\!1\!+\!\tau)^{\frac{1}{2}}}, 
\label{eq:disc-potential-alpha-0-coeff} 
\end{eqnarray} 
when $\alpha=0$. Here $R_F$ is the Carlson elliptic integral of the
first kind (e.g., Press et al.\ 1992), and we have corrected some
typographical errors in expressions (5.10) and (5.11) of Evans \& de
Zeeuw (1992).

For the specific case with $S(\bar \mu)$ given in eq.~(\ref{eq:s-mu}), 
the
integral equation (\ref{eq:integral-equation}) can be solved to give
(with $P=p$)
\begin{equation} 
w(\tau)=\alpha(1\!+\!p^2) \delta(\tau)+2p^2 \delta'(\tau), 
\end{equation} 
where $\delta(\tau)$ is the Dirac delta function and $\delta'(\tau)$
is its derivative (see, e.g., Morse \& Feshbach 1953, p.\ 837 or
Andrews \& Shivamoggi 1986, p.\ 35). The delta functions reduce the
integrals (\ref{eq:disc-potential-alpha-0-coeff}) and
(\ref{eq:disc-potential-alpha-not-0}) to single integrations. We find
\begin{eqnarray} 
V(R,\bar \mu)\!\!\!\!\! &=& \!\!\!\!\! {2\pi G \Sigma _0 p^{2+\alpha} 
R^{\alpha} 
\over \alpha {\rm B}\left ( \frac 12, \frac 12 \!-\!\frac {\alpha}2 
\right)} 
\int\limits_{0}^{\infty} \!\!\!\! 
      {(u\!+\!\bar \mu)^{\frac{\alpha}{2}}{\rm d}u  \over 
u^{\frac{1}{2}}[(1\!+\!u)(p^2\!+\!u)]^{\frac{1}{2}+\frac{\alpha}{2}}}  
                                                          \nonumber \\ 
\!\!\!\!\! &{}& \!\!\!\!\! \times \left ( {1\!+\!\alpha \over 
p^2\!+\!u}+ 
{1\!+\!\alpha \over 1\!+\!u} -{\alpha \over u\!+\!\bar \mu} \right ), 
\label{eq:ppd-potential-gen}
\end{eqnarray} 
for $\alpha \not = 0$. For $\alpha=0$ the coefficients $A$ and $B$ in
eq.~(\ref{eq:disc-potential-alpha-0}) are given by
\begin{eqnarray} 
A \!\!\!\!\! &=& \!\!\!\!\! \frac 43 p^2 G  
                    \Sigma _0 [ R_D(0,1,p^2)\!+\!R_D(0,p^2,1)], 
\nonumber \\ 
B(\bar \mu) \!\!\!\!\! &=& \!\!\!\!\! 2 p^2 G \Sigma _0 
\int\limits_{0}^{\infty} 
\left ( \frac 1{u\!+\!p^2}\! +\! \frac 1{u\!+\!1} \right) 
         \frac {\ln (u\!+\!\bar \mu){\rm d}u} 
{\sqrt{u(u\!+\!p^2)(u\!+\!1)}}    \nonumber \\ 
\!\!\!\!\! &{}& \!\!\!\!\! - \frac 43 p^2 G \Sigma _0 R_J(0,p^2,1,\bar 
\mu), 
\label{eq:ppd-potential-spec}
\end{eqnarray} 
where $R_D$ and $R_J$ are the Carlson elliptic integrals of the second
and third kind, available in the standard numerical libraries.
\end{document}